\documentclass[onecolumn,showpacs,floatfix,11pt,nofootinbib]{revtex4}
\usepackage{color}

\def\bq{\begin{quote}}
\def\eq{\end{quote}}

 at 10truept

\newcommand{\beq}{\begin{equation}}
\newcommand{\eeq}{\end{equation}}
\newcommand{\beqa}{\begin{eqnarray}}
\newcommand{\eeqa}{\end{eqnarray}}

\usepackage{amssymb,amsmath}
\usepackage{graphicx,float}
\usepackage{indentfirst}
\newcommand{\be}{\begin{equation}}
\newcommand{\ee}{\end{equation}}

\newcommand{\ba}{\begin{eqnarray}}
\newcommand{\ea}{\end{eqnarray}}

\begin{document}\title{Regular Bulk Solutions and Black Strings from Dynamical Braneworlds with Variable Tension}
\author{D. Bazeia}
\email{bazeia@fisica.ufpb.br}
\affiliation{Instituto de F\'\i sica, Universidade de S\~ao Paulo,  05314-970, S\~ao Paulo, SP, Brazil}
\affiliation{Departamento de F\'\i sica, Universidade Federal da
Para\'\i ba, 58051-970, Jo\~ao Pessoa, PB, Brazil}
\author{J. M. Hoff da Silva}
\email{hoff@feg.unesp.br}
\affiliation{Departamento de F\1sica e Qu\1mica, Universidade
Estadual Paulista, Av. Dr. Ariberto Pereira da Cunha, 333,
Guaratinguet\'a, SP, Brazil.}
\author{Rold\~ao da Rocha}
\email{roldao.rocha@ufabc.edu.br}
\affiliation{Centro de Matem\'atica, Computa\c c\~ao e Cogni\c c\~ao, Universidade Federal do ABC,  09210-170, Santo Andr\'e, SP, Brazil.}

\begin{abstract}
Regular black strings solutions associated to a dynamical Friedmann-Robertson-Walker  braneworld are obtained as a particular case of a regular bulk metric, in the context of a variable brane tension.  By analyzing the 5D Kretschmann invariants, we show that the variable brane tension is capable to remove bulk singularities, along some eras of the evolution of the Universe. In particular, the black string time-dependent profile in the bulk is analyzed in the context of the McVittie metric on an E\"otv\"os fluid braneworld, wherein the fluid dynamical brane tension depends on the brane temperature.  The whole bulk is shown to be regular for a Universe dominated by non-relativistic matter or  by relativistic matter/radiation, as the cosmological time elapses. In a particular case the associated black strings are shown to have finite extension along the extra dimension, what prominently  modify the higher-dimensional physical singularities. For a Universe dominated by matter or radiation,  we show that the brane singularities are removed along the extra dimension, as an effect of the variable brane tension, corresponding also to the vanishing of  the black string horizon. In this case there is a regular solution in the bulk.  
We show further that no additional physical singularity is introduced in the bulk when the Universe is dominated by a cosmological constant.

\end{abstract}\maketitle
\flushbottom

%\pacs{11.25.-w, 04.50.-h, 04.50.Gh}

\section{Introduction}

Braneworld models with variable brane tension pave attempts to probe signatures arising from high energy physics. Indeed, the drastic modification of the temperature of the Universe along its cosmological evolution instigates a  braneworld scenario with variable tension, generalizing the original Randall-Sundrum model \cite{ran},  in order to permit Friedmann branes \cite{maar2000, binetruy, bazeia1}. The variable tension brane dynamics was investigated in \cite{gly1, gly2, bulk1,bulk2}, also in a particular model, where the brane tension has an exponential dependence with the scale factor \cite{european}. 
Branes are dynamical objects emerging as topological defects in field theory or solitonic vacuum solutions in string theory. They can fluctuate and are described by their inherent degrees of freedom, as their tension. Otherwise the brane is rigid and there are no branons associated to it \cite{branons}.

The field equations regarding a brane containing any type of matter with a cosmological constant in the bulk
were introduced in \cite{binetruy}. There, the Friedmann equation in the context of a braneworld scenario was obtained,
as well as the time dependent scale factor. Solutions designating realistic black holes on the brane (at least stable and absent of naked singularities) are not straightforward to be obtained. Computational methods regarding relativistic stars on the brane and the exact solution of the collapse on the brane as well, in the AdS/CFT correspondence framework \cite{emp2,emp3}, are some endeavours to solve this question \cite{maartens, yoshino}. 
There are some arguments that whatever the solutions are, it may tend to a static geometry at late times, close to the Schwarzschild one, at large distances \cite{maartens}. %In this paper we regard
%a black hole geometry in a realistic braneworld scenario, provided by the McVittie metric.

Motivated by some generalizations concerning how black holes on braneworld scenarios capable to inquest the bulk metric \cite{Gergely:2006hd,Anderson:2005af},  black strings associated to a Friedmann-Robertson-Walker (FRW)  on the brane are here presented in a variable tension brane scenario.   
Our main aim here is to consider a Schwarzschild black hole embedded in a FRW brane \cite{mcvittie}. The McVittie metric is known to be regular everywhere and  the solutions asymptote in the future, and near the horizon, to the Schwarzschild-de Sitter geometry. Such description holds if the cosmological scenario is dominated at late times by a positive cosmological constant \cite{Kaloper:2010ec}.  For solutions without a positive cosmological constant the horizon is a {weak soft singularity}, but the metric can be extended, being possible to regard the causal character of the singularity.

McVittie solutions are spherically symmetric, parametrized by an scale factor $a(t)$  and a mass parameter $M$. They are led to black hole metrics or the standard FRW cosmology in suitable limits. These properties make these solutions  physically relevant for describing real gravitating objects in the Universe. 
These solution has been the
subject of a huge amount of investigations \cite{plb2013,lake1,Carrera:2009ve, faraoni,nolann,abd,lake2,
Ferraris:1996ey,Haines:1993sd,Patel:1999ej,sakaihaines}, and the misinformation around this solution was  only recently provided, what poses this solution as a true  black hole model \cite{Kaloper:2010ec, lake1}.  The spacetime related to the McVittie solution can be further generalized, providing an instrument to a black hole mass variation, subsequently influencing the associated black string. On the brane, this more general solution regards the description of some systems presenting a coupling between the local gravitational attraction  and  expansion of the 4D Universe \cite{abd}, described by the braneworld.

{In this paper we analyze the evolution of the black string warped horizon in a variable tension brane framework along the extra dimension \cite{meuhoff}, where the corrections in the black string warped horizon arise, in the three stages of the Universe evolution.
We shall show that on a variable tension braneworld, the associated black string may present finite extent along the extra dimension, even near the brane.
Such study has huge importance, in particular in what concerns the black string stability/instability, as analyzed originally for the (Schwarzschild)  black string, \`a la Gregory and Laflamme \cite{greg}.  In fact, when the Universe is dominated by non-relativistic matter or  by relativistic matter/radiation, the  black string extent is going to be shown here to be finite, and the bulk solution is an exact regular solution: the 5D singularities in the bulk are removed as the cosmological time elapses, due to the an immediate effect of the variable brane tension.  
Moreover, we investigate the physical singularities in the bulk, determined by the curvature invariants. When a  brane metric with a horizon is considered, it is an well established fact that it can generate in the bulk an additional  singularity at the location of
the horizon, and if a
brane metric with no horizon is used instead then no additional singularity
appears in the bulk \cite{Kanti1,Kanti2}. We shall reveal that feature regarding a black string associated to  the McVittie metric on the brane, by investigating the 4D and 5D Kretschmann invariants.}
Such two aims overlap and answer  relevant questions. In fact, as the singularity structure of the higher-dimensional spacetime is of great
importance in deciding whether a particular solution is physically acceptable, by analyzing the 5D Kretschmann invariant we show that  the black string warped horizon vanishes in some point along the extra dimension, for a Universe dominated by a cosmological constant and it makes the brane singularities to disappear. In other words, the vanishing of the warped horizon leaves behind a regular
bulk. We show further that in the eras corresponding to a Universe dominated by matter or radiation, the physical singularities in the brane remain in the bulk, where no further singularities are formed. 

Our program throughout this paper explicitly consists of the following: the next Section briefly revisits the way how the bulk metric along the extra dimension can be obtained, from exclusive information on the brane. It is accomplished by  studying the black string warped horizon along the extra dimension, and considering terms including the variable brane tension.  In Section 3 we delve into a black hole metric on a FRW  braneworld with variable tension, under the E\"otv\"os law. The associated black string warped horizon profile is studied for the three standard cases: the brane dominated by a) non-relativistic matter; b) radiation or relativistic matter; c) a cosmological constant. Each case is deeply analyzed in the context of a variable brane tension as well as their physical consequences. The case a) and b) present unexpected properties, namely, when the variable brane tension is taken into account 
there is an era beyond which the black string is warped horizon is zero, and  the black string ceases to exist. In fact, {in Section 4 the black string physical singularities are analyzed. We show that when the black string horizon goes to zero, it implies also that the singularities are banished in the bulk, what makes a regular bulk solution. For a Universe dominated by a cosmological constant, the physical singularities on the brane remain in the bulk, and no additional singularity
appears in the bulk. }

\section{Black String Warped Horizon and Variable Brane Tension}

In a braneworld with a single extra dimension of infinite extent, a vector field in the  bulk decomposes into components 
in the brane and orthogonal to the brane, as $ (x^{\alpha},y)$. The bulk is endowed with a
metric $\check{g}_{AB}dx^A dx^B = g_{\mu\nu}(x^\alpha,y)\,dx^\mu dx^\nu + dy^2$. The brane metric components $g_{\mu\nu}$ and the bulk metric are related by
$
\check{g}_{\mu\nu} = g_{\mu\nu} + n_\mu n_\nu, 
$ where the $n^\sigma$ are time-like vector field components, splitting the bulk. Moreover,  $g_{55} = 1$ and $g_{j5} = 0$, $\kappa^2_{4}=\frac{1}{6}\lambda\kappa^4_5$ and $
\Lambda_4=\frac{\kappa_5^{2}}{2}\Big(\Lambda_{5}+\frac{1}{6}\kappa_5^{2}\lambda^{2}\Big)$,
where $\Lambda_4$ is the effective brane cosmological constant,
$\kappa_4$ [$\kappa_5$] denotes the 4D [5D]
gravitational coupling, and $\lambda$ is the brane tension. The
extrinsic curvature {\ba
K_{\mu\nu}&=&-\frac{1}{2}\kappa_5^2 \left(T_{\mu\nu}+ \frac{1}{3}
\left(\lambda-T\right)g_{\mu\nu} \right).\label{kurv}
\ea}\noindent is obtained by using the  junction conditions. 
 Hereupon $T^{\mu\nu}$ is the energy-momentum tensor, for any 2-tensor $D_{\mu\nu}$, we shall adopt the convention $D=D_\mu^{\;\mu}$, and $D^2 = D_{\rho\sigma}D^{\rho\sigma}$.
One defines ${E}_{\mu\nu} = C_{\mu\nu\sigma\rho} n^\sigma n^\rho$ and  ${\cal B}_{\mu\nu\alpha} = g_\mu^{\;\rho} g_\nu^{\;\sigma}
C_{\rho\sigma\alpha\beta}n^\beta$ where $C_{\mu\nu\sigma\rho}$ is the 5D Weyl tensor.

The field equations together with the 5D Einstein and Bianchi equations \cite{3333,Gergely:2006hd, maartens} are used  to compute the bulk metric near the brane, and in particular the black string warped horizon along the extra dimension \cite{maar2000, casadio2004}. Such procedure provides  informations on all the bulk metric components \cite{meuhoff}, given by (denoting $g_{\mu\nu}(x^\alpha,0) = g_{\mu\nu}$):
 \ba
\hspace*{-0.6cm}{\;\;\;\;}g_{\mu\nu}(x^\alpha,y)&\!\!=\!\!& g_{\mu\nu}-\kappa_5^2\left(
T_{\mu\nu}\!+\!\frac{1}{3}(\lambda-T)g_{\mu\nu}\right)|y|\nonumber\\&+&\! \left[\frac{1}{2}\kappa_5^4\!\left(
T_{\mu\alpha}T^\alpha_\nu  +\frac{2}{3} (\lambda-T)T_{\mu\nu}
\right)\!-2{E}_{\mu\nu}+\left( \frac{1}{18}
\kappa_5^4(\lambda-T)^2-\frac{\Lambda_5}{3}
\right)\!g_{\mu\nu}\right] \frac{y^2}{2!} {\;}\nonumber\\
&+&\!\!\!\left.\Bigg[2K_{\mu\beta}K^{\beta}_{\alpha}K^{\alpha}_{\nu} - ({E}_{\mu\alpha}K^{\alpha}_{\nu}+K_{\mu\alpha}{E}^{\alpha}_{\nu})-\!\frac{1}{3}\Lambda_5K_{\mu\nu}\!-\!\nabla^\alpha{\cal B}_{\alpha(\mu\nu)} \!+ \!\frac{\Lambda_5}{6}\!\left(K_{\mu\nu}\!-\!g_{\mu\nu}K\right)
 \right.\nonumber\\
&+&\left.K^{\alpha\beta}R_{\mu\alpha\nu\beta} + 3K^\alpha{}_{(\mu}{\cal
E}_{\nu)\alpha}-K{E}_{\mu\nu}+\left(K_{\mu\alpha}K_{\nu\beta}
-K_{\alpha\beta}K_{\mu\nu}\right)K^{\alpha\beta}-\frac{\Lambda_5}{3}K_{\mu\nu}\Bigg]\;\frac{|y|^3}{3!}\right. \nonumber\\&+&\left.\Bigg[\frac{\Lambda_5}{6}\left(R-\frac{\Lambda_5}{3} + K^2\right)g_{\mu\nu} + \left(\frac{K^2}{3}- \Lambda_5\right)K_{\mu\alpha}K^{\alpha}_{\;\nu} + (R-\Lambda_5 + 2K^2){E}_{\mu\nu}\right.\nonumber\\&-&\left. \frac{13}{2}K_{\mu\beta}{E}^\beta_{\;\alpha}K^{\alpha}_{\;\nu}+\left(K^{\alpha}_{\;\sigma}K^{\sigma\beta} \!+ {E}^{\alpha\beta}\! +KK^{\alpha\beta}\right)R_{\mu\alpha\nu\beta} - \frac{\Lambda_5}{6}R_{\mu\nu}
 + 2 K_{\mu\beta}K^{\beta}_{\;\sigma}K^\sigma_{\;\alpha}K^\alpha_{\;\nu}\right.\nonumber\\&+&\!\!\left.   {E}_{\mu\alpha}\!\left(\!K_{\nu\beta}K^{\alpha\beta}\!-3K^\alpha_{\;\sigma}K^{\sigma}_{\nu}\! +\! \frac{1}{2}KK^\alpha_{\nu}\!\right)+K_{\sigma\rho}K^{\sigma\rho}K\,K_{\mu\nu}-\! K_{\mu\alpha}K_{\nu\beta}{E}^{\alpha\beta}\right.\nonumber\\
&+&\left.
 \left(\frac{7}{2}KK^{\;\alpha}_{\mu}- 3K^\alpha_{\;\sigma}K^\sigma_{\;\mu}\right)\!{E}_{\nu\alpha}+\left(3K^\alpha_{\mu}K^{\beta}_{\alpha}\!-\!K_{\mu\alpha}K^{\alpha\beta}\right)\!{E}_{\nu\beta}\!\! -4K^{\alpha\beta}R_{\mu\nu\gamma\alpha}K^{\gamma}_{\;\beta}\!\right.\nonumber\\&-&\left.\!  \! \frac{7}{6}K^{\sigma\beta}K^{\;\alpha}_{\mu}R_{\nu\sigma\alpha\beta}\Bigg]\,\frac{y^4}{4!} + \cdots \right.  \label{eletrico} \ea
The black string warped horizon   $\sqrt{g_{\theta\theta}(x^\alpha,y)}$ \cite{clark} is given by (\ref{eletrico}) for $\mu=\theta=\nu$.
This Taylor expansion is not a perturbation and the brane is not bent, providing thus the continuity of $g_{\mu\nu}(x^\alpha,y)$  when there are discontinuities in the matter stress tensor or in the the tensors $E_{\mu\nu}$ and ${\cal B}_{\mu\nu\sigma}$  \cite{casadio2004}. This regularity can be viewed further  in a microscopic description of the braneworld: matter should be smooth into the bulk, and further  localized on the brane within a width  $\sim\lambda^{-1/2}$ \cite{all}. 

In a variable brane tension scenario, the terms in $|y|$ and $y^2/2!$ in (\ref{eletrico}) present no additional terms. Notwithstanding, starting from the order $|y|^3/3!$, those additional terms play a fundamental role. The term that contributes to the derivatives of the variable tension $\lambda$ in the order $|y|^3/3!$ in (\ref{eletrico}) reads:
\ba
-\frac{2\kappa_5^2}{3}\left((\nabla^\alpha\nabla_\alpha\lambda)g_{\mu\nu}-(\nabla_{(\nu}\nabla_{\mu)}\lambda)
\right).\label{magnetico}
\ea
Terms in order $y^4/4!$ are obtained arising from the variable tension brane in 
(\ref{eletrico}) are given by  
\ba
&&6 \left[(\Box\lambda)K_{(\mu\tau}{E}_{\nu)}^\tau - \nabla^\alpha((\nabla_{(\mu}\lambda)\, {E}_{\nu)\alpha})\right]
+2\left(K + \frac{7}{3}\kappa_5^2\right)\left[(\Box\lambda)K\,K_{\mu\nu}-\nabla^\alpha((\nabla_{(\mu}\lambda)\, K\,K_{\alpha\vert\nu)})\right]\nonumber\\
&&- \frac{1}{3}\kappa_5^2\,\left[\Box(\Box\lambda)g_{\mu\nu}-\nabla_{(\nu}\nabla_{\mu)}(\Box\lambda)\right]+\left(\frac{1}{3}\kappa_5^2+2 K\right)[(\Box\lambda){E}_{(\mu\nu)} - \nabla^\alpha\left((\nabla_{(\mu}\lambda)\, {E}_{\nu)\alpha}\right)]\nonumber\\
&&+ \frac{1}{3}\kappa_5^2 \left[(\Box\lambda)
(R_{\mu\nu}+K_{(\mu\vert\tau}K_{\nu)\beta} 
K^{\tau\beta} - K^2\,K_{(\mu\nu)})-
\nabla^\alpha((\nabla_{(\mu}\lambda)
(R_{\alpha\vert\nu)}  - 
K_{\alpha\tau}K^{\tau}_{\nu)} - K
\,K_{\alpha\nu)}))\right]
\nonumber\\&&
-2 K^{\tau\beta}\left[(\Box\lambda)R_{(\mu\vert\tau\vert\nu)\beta} - \nabla^\alpha\left((\nabla_{(\mu}\lambda)\, R_{\alpha\tau\vert\nu)\beta}\right)\right]
+\left(2\,K^2-\frac{1}{3}\Lambda_5 \right)[(\Box\lambda)g_{\mu\nu}-\nabla_{(\nu}\nabla_{\mu)}\lambda].\label{magnetico1}
 \ea
A   time-dependent brane tension $\lambda = \lambda(t)$ is going to  be considered, what turns the expressions (\ref{magnetico}, \ref{magnetico1}) to a simpler form.

\section{Bulk Metric and Black String from a Dynamical  E\"otv\"os Braneworld}

This Section is devoted to derive bulk metric solutions near the brane, and in particular the black string profile, associated to a FRW E\"otv\"os variable tension braneworld.

The McVittie solution in isotropic spherical coordinates {\rm r} defined implicitly by by $ {r} = {\rm r}\left(1+ \frac{2GM}{{\rm r}}\right)^2$, as in \cite{mcvittie, buchdahl,Kaloper:2010ec} reads 
\be g_{\mu\nu}dx^\mu dx^\nu = - \Bigl(\frac{1-\mu}{1+\mu}\Bigr)^2 dt^2 + (1+\mu)^4
a^2(t) (d{\rm r}^2 +{\rm r}^2d\Omega^2)\, , \label{mcvitt} \ee
where $a(t)$ is the cosmological scale factor and one denotes $\mu= \frac{M}{2a(t) r}$. 
The metric (\ref{mcvitt}) is an exact solution of the Einstein field equations when the scale factor  solves the
Friedmann equation 
\be  \rho(t) = \frac{3\dot{a}^2}{8 \pi Ga^2}. \label{friedmann} \ee
As usual, $\rho$ denotes the energy density and $H(t) = \dot a(t)/a(t)$ denotes the Hubble parameter.  The pressure is given by \cite{sakaihaines}
\be
p = -\frac{1}{8\pi G} \left(3 \frac{\dot{a}^2}{a^2} +\frac{2}{\beta} \left(\frac{\ddot{a}}{a}-\frac{\dot{a}^2}{a^2}\right)  \right), 
\label{pr1}
\ee\noindent where $\beta:= \frac{1-\mu}{1+\mu}$.
The McVittie solution is shown to be  the unique solution describing the field of a spherically symmetric mass in a spatially flat asymptotically FRW cosmology  \cite{nolann}. Notwithstanding, due to stringent assumptions, non-gravitational forces encrypted by the inhomogeneous pressure (\ref{pr1}) are demanded.

In order to analyze the black string 
associated to the McVittie metric,  we focus on the term $g_{\theta\theta}(x^\alpha,y)$  of the expansion at Eq.~(\ref{eletrico}), as it represents the bulk metric near the brane and, in particular, the  square of the black string warped horizon along the extra dimension, when $\mu = \nu = \theta$. Clearly, a time dependent brane tension modifies the black string
 associated to the McVittie solution. 
The projected Weyl term on the brane is given by \cite{maartens}
\begin{eqnarray}
\!\!\!\!\!{E}_{\theta\theta}(r,t)&=& \frac{1}{4(1+\mu)^4a^2}\! +\frac{\rho^2}{3\beta^4}\left(1-2\beta\right)^2-\! \frac{\rho p}{4a^2}\left(4\mu^3+5\mu^2+4\mu+1\right)\,.
\label{weyl}
\end{eqnarray} 
{In addition, the extrinsic curvature in (\ref{kurv}) can be expressed as
\ba
K_{\theta\theta}=\kappa_5^2\left(3\frac{\dot{a}^2}{a^2}+\frac{\lambda}{3}+\frac{\dot{H}}{\beta}+\frac{3\dot{a}^2}{2\beta^6}\right)\,.\label{kurvmc}
\ea}
\noindent 
By substituting the expressions (\ref{friedmann}, \ref{pr1}), the black string warped horizon $g_{\theta\theta}(x^\alpha,y)$ in (\ref{eletrico}) is
 \ba
g_{\theta\theta}({\rm r},t,y)&=&{\rm r}^2\left[1-\kappa_5^2\left(3\frac{\dot{a}^2}{a^2}+\frac{\lambda}{3}+\frac{\dot{H}}{\beta}+\frac{3\dot{a}^2}{2\beta^6}\right)|y|\right]\,\nonumber\\
&+&{\rm r}^2\left[\!\frac{3\dot{a}^4}{16 a^4}\!\!\left[\frac{\left(2\beta- 1\right)^2}{6\beta^4}+\frac{9\beta^2}{4}-\frac{3\beta}{2}\!+\!\frac{145}{4(1+\mu)^4a^2}\!+\! \left(\frac{3\dot{a}^2}{a^2}+\!\frac{2\dot{H}}{\beta}\right)\!
\left(\frac{4\mu^2+\mu+3}{(\mu-1)(1+\mu)^4}\right)\right]\right.\nonumber\\
&+&\left.\! \!\!\frac{\kappa_5^4}{36}\!\left(\!\lambda \!+\! \frac{3\dot{a}^2}{2\beta^2}(1\!+\!\mu)^4\!+ \!\frac{9\dot{a}^2}{2} \!+ \! \frac{3\dot{H}^2}{\beta}\!\right) \!+\!\frac{3\beta}{2(1\!+\!\mu)^4a^2}\!+\! \frac{2\lambda}{3} \!-\!
 \frac{\Lambda_5}{6}(1\!+\!\mu)^4 a^4\right]\!\!\frac{y^2}{2!} + \cdots
\label{mag2}
 \ea\noindent where $\dot{H}= \!\left(\frac{\ddot{a}}{a}\!-\!\frac{\dot{a}^2}{a^2}\right)$. 
Eq.~(\ref{mag2}) is written here up to the second order in the extra dimension for the sake of conciseness, although the awkward expansion including the term $y^4/4!$ was considered in \cite{meuhoff}, and shall be adopted in the graphics in the end of this Section.  

When  $a(t)=1$, by transforming the spherical isotropic coordinates to the spherical standard ones, the bulk metric component in (\ref{mag2})   is led to  \begin{eqnarray}
g_{\theta\theta}({\rm r},y)&=&{\rm r}^2\left[1-\frac{\kappa_5^2\lambda}{3}\,|y|~~{}+\frac{1}{3}\left(\frac{1}{6}\kappa_5^4\lambda^2 - \Lambda_5\right)\, \frac{y^2}{2!}-\left(\frac{193}{216}\lambda^3\kappa_5^6 +\frac{5}{18}\Lambda_5\kappa_5^2\lambda\right)\,\frac{|y|^3}{3!}  \right.\nonumber\\&&\qquad\qquad\qquad\left.-\frac{1}{18}\Lambda_5\left(\Lambda_5 + \frac{1}{6}\lambda^2\kappa_5^4+\frac{7}{324}\lambda^4\kappa_5^8\right)\,\frac{y^4}{4!} +
\cdots\right],\end{eqnarray} which is the classical Schwarzschild black string warped horizon when {\rm r} corresponds to the coordinate singularity \cite{plb2013,Chamblin:1999by, maartens, clark, meuhoff} is obtained in this limit. Indeed, the physical content regarding  the black string in (\ref{mag2}) is based upon the analysis of an invariant \cite{Kaloper:2010ec}. When  $a(t)=1$, the associated Kretschmann scalars are led to the Schwarzschild ones.  

With respect to Eq.(\ref{magnetico1}) 
 for the McVittie metric, Eq.(\ref{magnetico}) has a straightforward form given by {\begin{eqnarray}\label{ele1}-\frac{2\kappa_5^2\lambda''}{3},\end{eqnarray}\noindent} and in  order $y^4/4!$, are given by:
 \begin{eqnarray}\label{ele2}
&&\!\!\!\frac{16 a^6 {\rm r}^4 \left(\dot{a}^2 (2 a {\rm r}-5 M)+a \left(2 \ddot{a}^2 (2 a {\rm r}+M)+a \lambda  (2 a
   {\rm r}-M)\right)\right)}{(M-2 a {\rm r}) (2 a {\rm r}+M)^4}+\frac{6 \left(\dot{a}^2-a \ddot{a}^2\right) (2 a {\rm r}+M)}{2 a {\rm r}-M}\!\!\nonumber\\
   &&\!\!\!+\frac{14 \kappa_5^2+5 \Lambda }{3}-9
   \dot{a}^2+6a^2+ \frac{2}{3 a^6{\rm r}^4 \left(1 + \frac{M}{2 a {\rm r}}\right)^8} \left(\frac{a^4 \left(\lambda -\frac{3 \dot{a}^2}{a^2}\right)}{\left(\frac{M}{2 a {\rm r}}+1\right)^8}+\frac{6
   \left(\dot{a}^2-a \ddot{a}^2\right) (2 a {\rm r}+M)}{2 a {\rm r}-M}-9 \dot{a}^2\right)\nonumber\\&&+\frac{a^4 \left(\lambda -\frac{3
   \dot{a}^2}{a^2}\right)}{\left(\frac{M}{2 a {\rm r}}+1\right)^8}+\frac{48 a^{10} {\rm r}^4}{(M-2 a {\rm r})^4} \left(\frac{M}{2 a
   {\rm r}}+1\right)^{12} \left(\frac{a^2 \left(\lambda -\frac{3 \dot{a}^2}{a^2}\right)}{3 \left(\frac{M}{2 a
   {\rm r}}+1\right)^8}+\frac{3 \dot{a}^2}{a^2}\right)+\frac{6 \left(\dot{a}^2-a \ddot{a}^2\right) (2 a {\rm r}+M)}{2 a
   {\rm r}-M}\nonumber\\&&+128 a^5 M^2 {\rm r}^5 \left(\left(6 a^2+2\right) {\rm r}^2\right)-9 \dot{a}^2+256 a^6 M {\rm r}^6 a^2
   {\rm r}^2\nonumber\\&&\frac{8 a \kappa_5^2 \lambda''}{3 {\rm r} (M-2 a {\rm r})^4 (2 a {\rm r}+M)^6}\left(\left(64 a^2 M^7 {\rm r}^2\!-\!512 a^7 {\rm r}^7\!+\!8 a M^6 {\rm r} \left(a^2
   {\rm r}^2\!-\!1\right)\!+\!16 a^2 M^5 {\rm r}^2 \left(\left(30 a^2-8\right) {\rm r}^2\!+\!7\right)\right.\right.\nonumber\\&&\left.\left.+128 a^4 M^3 {\rm r}^4 \left(\left(8
   a^2-4\right) {\rm r}^2+13\right)+64 a^3 M^4 {\rm r}^5 \left(13 a^2+6\right) {\rm r}^2+12 a M^8
   {\rm r}+M^9\right)\right)\nonumber\\
   &&\frac{\kappa_5^8 \lambda ''}{147456 a^{20} (2 a {\rm r}-M)} \left(4096 a^{18} \left(\dot{a}^2 (2 a {\rm r}-5 M)+a \left(2 \ddot{a}^2 (2 a {\rm r}+M)+a \lambda  (2 a
   {\rm r}-M)\right)\right)\right.\nonumber\\&&\left.-\frac{3 (2 a {\rm r}+M)^{12} \left(\dot{a}^2 (2 a {\rm r}-5 M)+2 a \ddot{a}^2 (2 a
   {\rm r}+M)\right)}{{\rm r}^{12}}\right)-\frac{\kappa_5^2 \lambda ^{(4)} (2 a {\rm r}+M)^4}{48 a^2 {\rm r}^4}\nonumber\\&&+\frac{\kappa_5^2 \lambda''}{2}\!\! \left(\frac{16 a^2
   {\rm r}^4}{(2 a {\rm r}+M)^4}\!+\!\frac{3 \dot{a}^2 \left(4 a^3 {\rm r}^3\!+\!8 a^2 M {\rm r}^2\!+\!5 a M^2 {\rm r}+2 M^3\right) \left(\dot{a}^2 (2 a {\rm r}-5
   M)+2 a \ddot{a}^2 (2 a {\rm r}+M)\right)}{4 a^7 {\rm r}^3 a^2 (2 a {\rm r}-M)}\right.\nonumber\\&&\left.\!+\!\frac{6 \dot{a}^4 (3 M\!-\!2 a {\rm r})^2}{a^4 (M-2 a
   {\rm r})^2}\right)\times\nonumber\\&&  \left(\frac{16 a^4 {\rm r}^4 \left(\dot{a}^2 (2 a\! {\rm r}\!-\!5 M)\!+\!a\! \left(2 \ddot{a}^2 (2 a {\rm r}\!+\!M)+\!a\! \lambda
    (2 a {\rm r}\!-\!M)\right)\right)}{(M-2 a {\rm r}) (2 a {\rm r}+M)^4}\!+\!\frac{3 \left(\dot{a}^2 (2 a {\rm r}-5 M)\!+\!2 a \ddot{a}^2 (2 a
   {\rm r}+M)\right)}{a^2 (2 a {\rm r}-M)}\right.\nonumber\\&&\left.+\frac{(2 a {\rm r}+M)^2 \left(\lambda  a^2\!-\!3 \dot{a}^2\right)}{3 a^2 (M\!-\!2 a
   {\rm r})^2}\!-\!\frac{3 \dot{a}^2}{a^2}\right)\!+\!4 M^2 \!\left(480 a^3 M {\rm r}^3\!+\!360 a^2 M^2 {\rm r}^2\!+\!16 a^6 {\rm r}^4   \lambda\!+\!120 a M^3 {\rm r}\!+\!15 M^4\right)
\nonumber\\&&+\frac{16 \kappa_5^2 \left(\lambda ''\right)^2 (M-2 a {\rm r})^3 \left(4 a^2 {\rm r}^2-2 a M {\rm r}+M^2\right)}{3 a^5 {\rm r}^3 \left(\frac{M}{2 a {\rm r}}+1\right)^{14}
   (M-2 a {\rm r})^4} \left(\left(2
   a \left(3 \ddot{a}^2 (2 a {\rm r}+M)^5\!-\!64 {}  a^7 (\lambda \!-\!5) {\rm r}^4 (2 a {\rm r}\!-\!M)\right)\right.\right.\nonumber\\&&\left.\left.-3 \dot{a}^2 (5 M-2 a {\rm r}) (2 a
   {\rm r}+M)^4\right)\right)-32 a M {\rm r}^7 (M-2 a {\rm r})^3 \left(4 a^2 {\rm r}^2-8 a M {\rm r}+M^2\right)\times\nonumber\\&&
   \left(2 a \left(64 {}  a^7 (\lambda -5) {\rm r}^4 (2 a {\rm r}-M)-3 \ddot{a}^2 (2 a {\rm r}+M)^5\right)+3 \dot{a}^2 (5 M-2 a {\rm r}) (2 a
   {\rm r}+M)^4\right)(2 a {\rm r}+M)^{12}\,.\nonumber\end{eqnarray}

We  assume the brane tension as an intrinsic property of the brane, just as an effective model \cite{gly1,gly2,bulk1,bulk2,european}. The brane tension is also supposed to be smooth, to have an inferior limit, and the brane tension fluctuations are  evanescent, in the sense that they  are suppressed exponentially. 

The huge variation of the temperature of the Universe in expansion needs to be modelled by a variable tension in the braneworld cosmology framework. 
The phenomenological E\"otv\"os law regarding standard fluid membranes \cite{Eotvos} is therefore used. Essentially, the E\"otv\"os law asserts that the (fluid) membrane tension depends on the temperature as \be \lambda=\chi (T_{c}-T),\label{TERM0}\ee where $\chi$ is a constant and $T_{c}$ is a critical temperature above what the membrane does not exist. The tension variation is now expressed in terms of the (cosmic) time, instead of the temperature. Indeed,  as the Universe expands, it cools down, and a variation on the temperature is nothing but a time variation. In the absence of stresses in the bulk there is no exchange of energy and momentum between the brane and the bulk \cite{maartens}. As  $dQ=dE+pdV=0$, by taking into account photons from the cosmic microwave background, it is possible to use the standard quantities $E=\sigma T^4V$ and $p={E}/{3}$, what implies that $\frac{dT}{T}=-\frac{1}{3}\frac{dV}{V}$. Finally, by expressing the volume in terms of the FRW scale factor ($V(t)=a^3(t)$) it implies that $T(t)\propto 1/a(t)$.   
This approach is in full agreement with the standard cosmological model \cite{meuhoff}. From  Eq.~(\ref{TERM0}) one obtains \be \lambda(t)=1-\frac{1}{a(t)}, \label{opa11} \ee where we normalize the brane tension and the scale factor as well.    

By delving into the analysis on the influence of the  brane tension variation on the black string associated to the McVittie solution,  two points support our effective approach. From the theoretical point of view, Eq.~(\ref{opa11}) is useful to merge supersymmetry and inflationary cosmology, since it engenders a time variable 4D cosmological constant,  which starts from a negative value and converges to a small positive one, as the universe expands \cite{gly2}. On the another hand, the type of time variation in  Eq.~(\ref{opa11}) is appropriate, from the experimental point of view. The projection scheme of bulk gravitational quantities \cite{3333}   evinces a linear dependence between the effective Newtonian constant and the brane tension, namely, $G\sim \lambda$. Hence, a time variation on the brane tension means a time variable gravitational constant. The best model independent bound on the fractional variation of $G$ is provided by the lunar laser ranging \cite{WILL}, which stays that $\dot{G}/G<(4.9\pm 5.7)\times 10^{-13} yr^{-1}$. When the expression (\ref{opa11}) is taken into account, the following bound  \be \frac{\dot{\lambda}(t)}{\lambda(t)}=\frac{\dot{a}(t)}{a(t)[a(t)-1)]},\label{novaaa} \ee is obtained, and all the inputs analyzed in this paper lead to a fractional variation of the brane tension satisfying the lunar laser ranging experiment, via Eq.~(\ref{novaaa}), for late times.

Regarding the McVittie solution, the Einstein field equations on the brane provide $
\rho \propto a^{-3(1+w)/\beta}$, where it is standard to define the state parameter $w = \frac{p}{\rho}$. It  leads to the time evolution of the scale factor. When $M=0$ it implies that $\beta = 1$, and the scale factor takes the well known value for the scale factor of a flat universe $a(t)\propto t^{{2}/{3}}$ (dominated by non-relativistic matter, where $w=0$) or $a(t)\propto t^{{1}/{2}}$ (dominated by the radiation or relativistic matter, where $w=\frac{1}{3}$). In the case of a cosmological constant ($w=-1$), it implies that $a(t)\propto \exp(H_0 t)$, independently of the value for $\beta$. The manifold $\mu=1$, corresponding to $\beta=0$, is the event horizon in the Schwarzschild case. Notwithstanding, one must avoid this value in order to circumvent the big bang singularity \cite{Kaloper:2010ec}. 

We shall compare the McVittie black string profile in the two eras of evolution of our Universe (without a cosmological constant) and, in addition, in the presence of a cosmological constant.
The pure cosmological constant braneworld scenario is displayed in Figs. 1 and  2, 
approaching a realistic black string in a global asymptotically FRW braneworld,
where locally the behavior of a solution is analyzed. In order to evince the difference between constant brane tension and variable  brane tension scenarios, in Fig. 1 we plot the McVittie black string for a constant brane tension analogue to  \cite{plb2013}, and in the graphics in Fig. 2  for the variable tension expanding braneworld.  The scenario is dominated by a cosmological constant  as the scale factor is given by $a(t) \propto \exp(H_0t)$.
According to (\ref{opa11}), the brane tension is given by \beq
\label{tensioneh0t}\lambda(t) = 1 - \exp(-H_0 t).\eeq \noindent
Hereupon in the graphics it shall be adopted $\Lambda_5 =\kappa_5=1$.
\begin{figure}[H]
\begin{center}\includegraphics[width=2.7in]{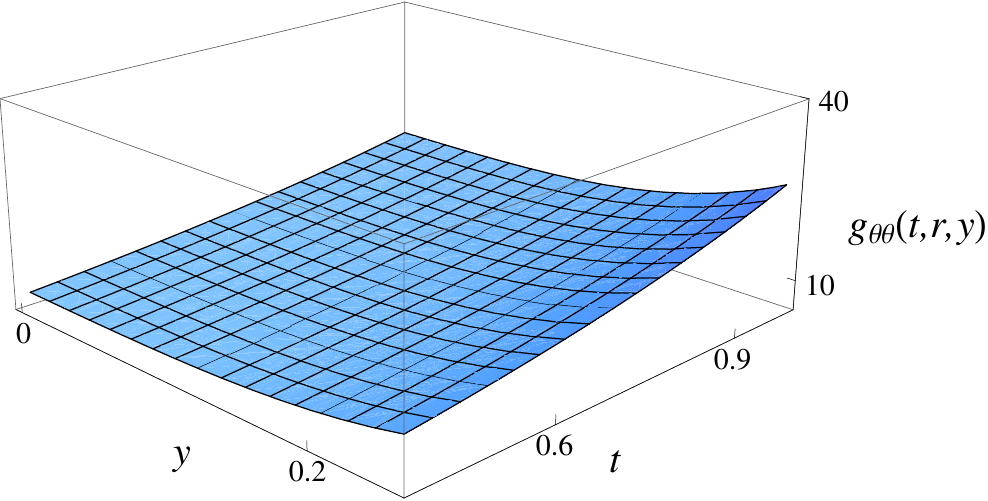}
\caption{\footnotesize\; Graphic of the brane effect-corrected warped   black string horizon $g_{\theta\theta}(t,{\rm r},y)$ associated to the McVittie solution on the brane with \emph{constant} tension, along the extra dimension $y$ and also as function of the time $t$. The scale factor $a(t) = \exp(H_0 t)$ is used. }
\end{center}
\end{figure}
\begin{figure}[H]
\begin{center}\includegraphics[width=2.7in]{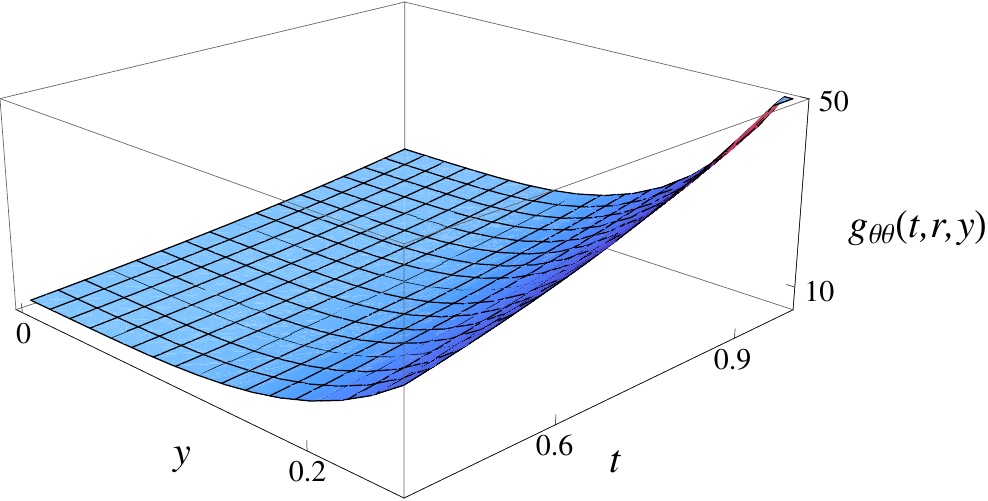}\quad\quad\includegraphics[width=2.7in]{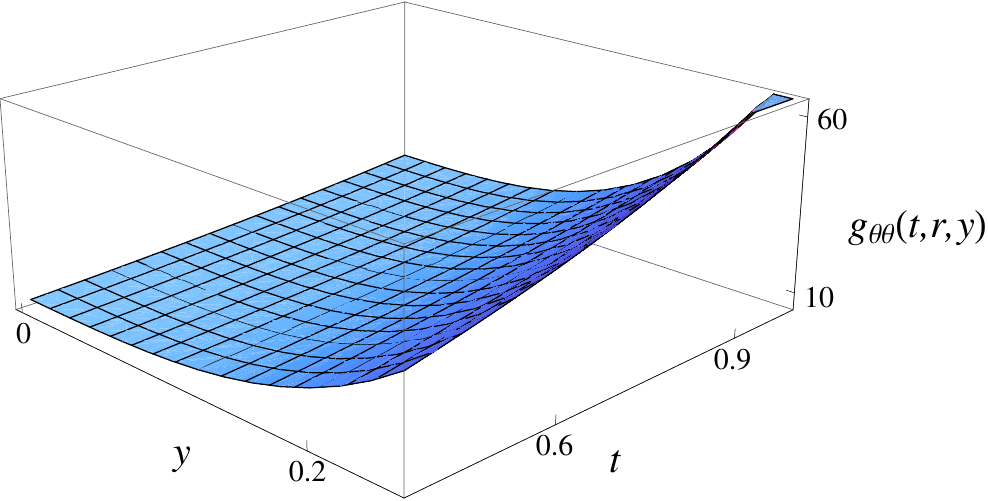}
\caption{\footnotesize\; Graphic of the brane effect-corrected warped   black string horizon $g_{\theta\theta}(t,{\rm r},y)$ associated to the McVittie solution on the brane with variable tension, along the extra dimension $y$ and also as function of the time $t$. These graphics respectively \emph{does not} and \emph{does} take into account the extra terms given by Eqs.~(\ref{magnetico}) and (\ref{magnetico1}), for the McVittie black string.}
\end{center}
\end{figure}

Now, the case where the scale factor $a(t) \propto t^{\beta/2}$ is considered, what emulates a a brane dominated by radiation. Therefore  Eq.~(\ref{opa11}) is written as
\be\label{tensiont2b} \lambda(t) = 1 - t^{-\beta/2}.\ee
As the situation previously analyzed, we first depict the black string  warped horizon for the brane tension $\lambda$ constant \cite{plb2013}, for the sake of ulterior comparison to the case of a variable brane tension: 
\begin{figure}[H]
\begin{center}\includegraphics[width=2.7in]{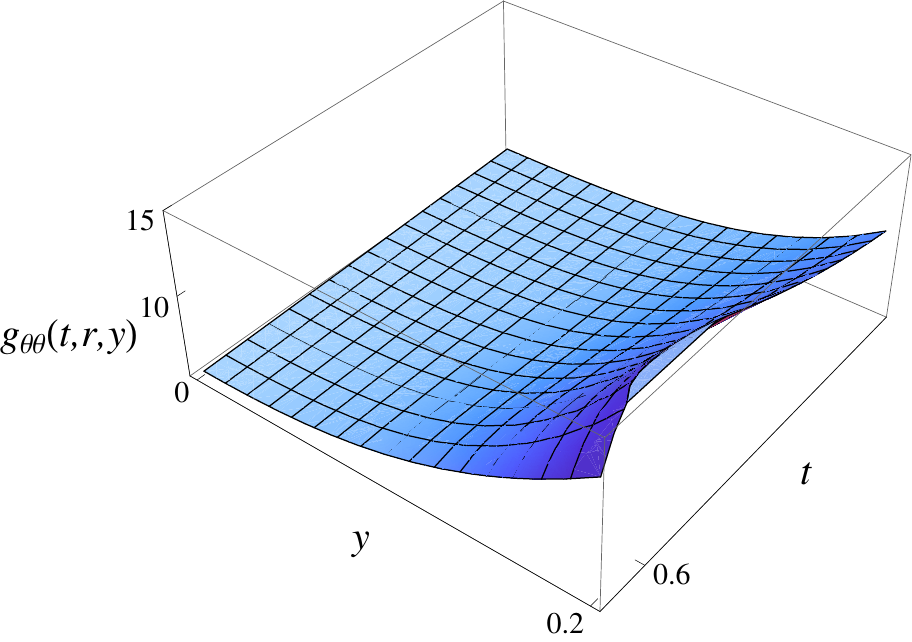}
\caption{\footnotesize\; Graphic of the black string warped horizon $g_{\theta\theta}(t,{\rm r},y)$ along the extra dimension $y$, as an explicit function of time $t$,  where $a(t) \propto t^{-\beta/2}$, for the McVittie metric.  }
\end{center}
\end{figure}
\begin{figure}[H]
\begin{center}\includegraphics[width=2.7in]{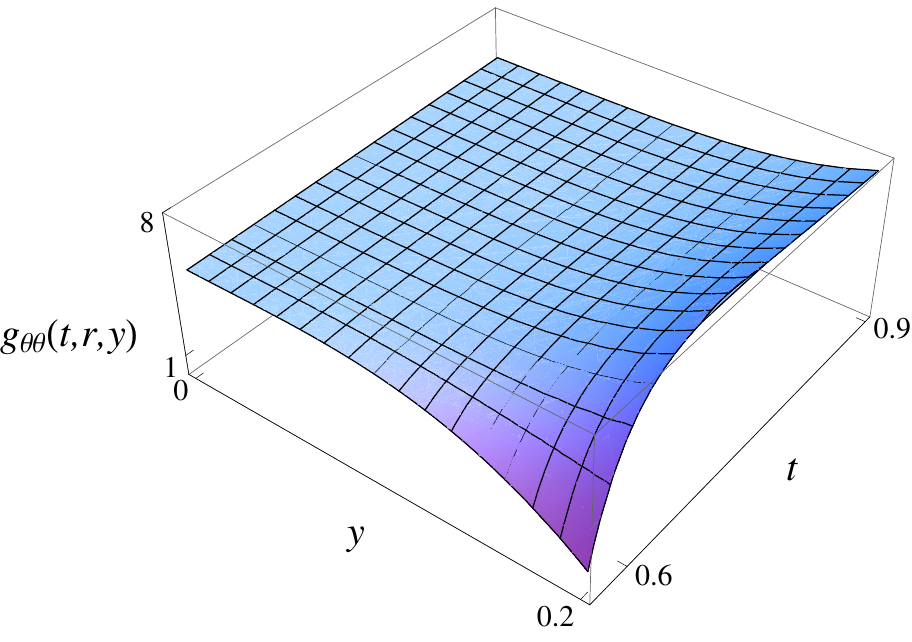}\quad\quad\includegraphics[width=2.7in]{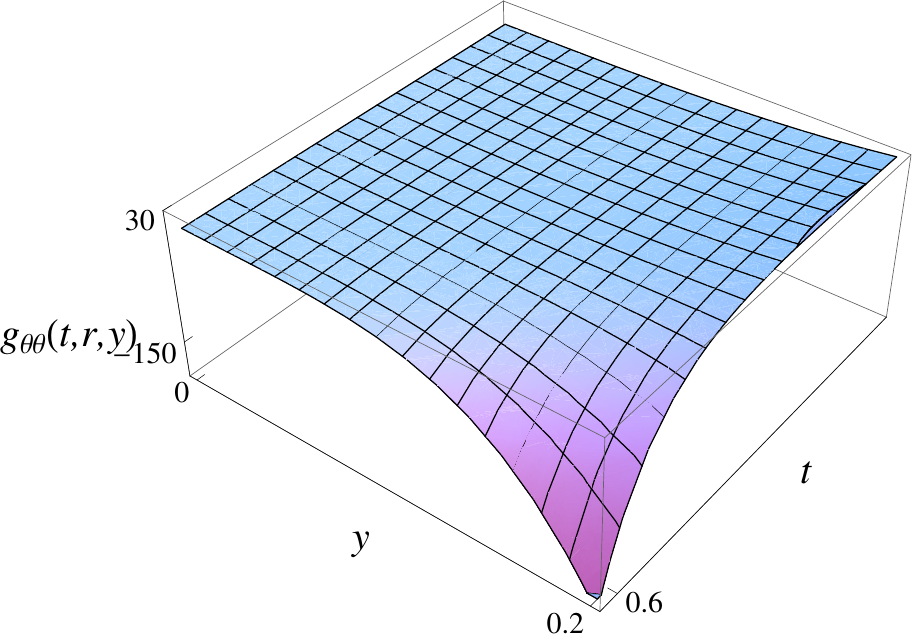}
\caption{\footnotesize\; Plot of the warped horizon $g_{\theta\theta}(t,{\rm r},y)$ along the extra dimension $y$, as an explicit function of time $t$, for $a(t) \propto t^{\beta/2}$. The brane tension is given by $\lambda(t) = 1 - t^{-\beta/2}$. These graphics respectively  \emph{does not} and \emph{does} take into account the extra terms given by Eqs.~(\ref{magnetico}) and (\ref{magnetico1}), for the McVittie black string.}
\end{center}
\end{figure}

In Fig. 1,  the
black 
string warped horizon 
for a brane dominated by a cosmological constant is illustrated, considering a constant brane tension. The graphics in  
Fig. 2 show respectively that a) the black string warped horizon profile  
different for a variable brane tension in such scenario, provided by Eq.~(\ref{tensioneh0t}); b) the black string, associated to the McVittie metric on the brane, has a different warped  horizon profile when  the extra terms given by
Eqs.~(\ref{magnetico}) and (\ref{magnetico1})  are taken into
account. Those
graphics show the paramount importance of considering more terms
in the metric expansion given by Eq.~(\ref{eletrico}), as accomplished
heretofore.  

By  concerning a brane dominated by radiation or relativistic matter $(a(t) \propto t^{\beta/2})$, as the time elapses, the black string warped horizon of the associated black string decreases along the extra dimension for any value for $t\lesssim0.53$ in Fig. 4, for a constant brane tension. For the time parameter greater than this value, the warped horizon of the black string always increases, along the extra dimension. Instead, the graphics in Fig. 5 take into account 
the variable brane tension in Eq.~(\ref{tensiont2b}), respectively without and with the extra terms arising from the variable brane tension in Eqs.~(\ref{magnetico}) and (\ref{magnetico1}). 

Now, the case where the scale factor $a(t) \propto t^{2\beta/3}$ emulates a matter-dominated brane is taken into account. In this case Eq.~(\ref{opa11}) is written as
\be\label{tension32b} \lambda(t) = 1 - t^{-2\beta/3}.\ee

\begin{figure}[H]
\begin{center}\includegraphics[width=2.7in]{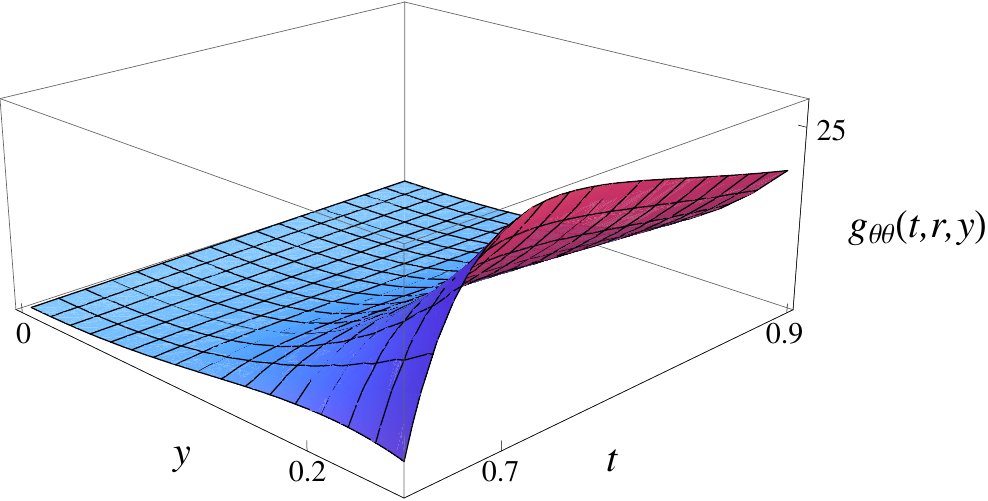}
\caption{\footnotesize\; Graphic of the warped horizon $g_{\theta\theta}(t,{\rm r},y)$ along the extra dimension $y$, as an explicit function of time $t$,  for $a(t) \propto t^{2\beta/3}$ for the McVittie metric.  }
\end{center}
\end{figure}
\begin{figure}[H]
\begin{center}\includegraphics[width=2.7in]{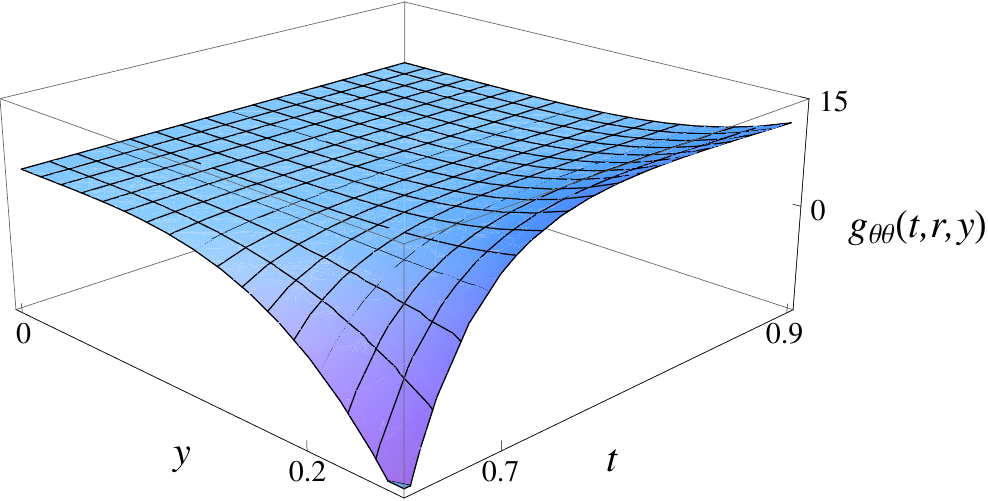}\includegraphics[width=2.7in]{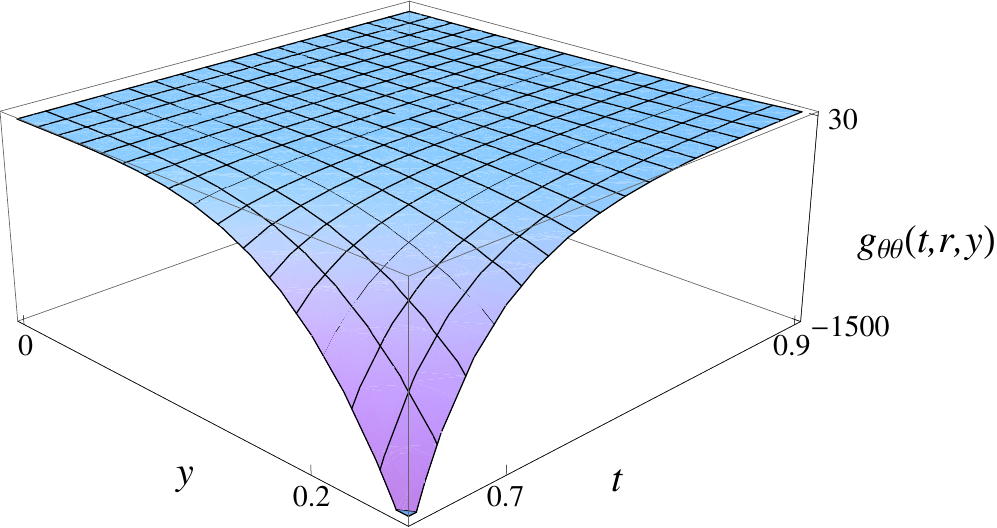}
\caption{\small  Graphic of the warped horizon $g_{\theta\theta}(t,{\rm r},y)$ along the extra dimension $y$, as an explicit function of time $t$, for $a(t) \propto t^{2\beta/3}$. The brane tension is given by $\lambda(t) = 1 - t^{-2\beta/3}$.  These graphics respectively  \emph{does not} and \emph{does} take into account the extra terms given by Eqs.~(\ref{magnetico}) and (\ref{magnetico1}), for the McVittie black string. The black string warped horizon decreases along the extra dimension $y$ for any value for $t\lesssim0.76$.
}
\end{center}
\end{figure}

The graphics on the left in Figs. 2, 4, and 6 do not take into account the time
derivative terms provided by Eqs.~(\ref{magnetico}) and (\ref{magnetico1}) respectively for each case analyzed, given by  (\ref{tensioneh0t}), (\ref{tensiont2b}), and  (\ref{tension32b}), while the graphics on the right in Figs. 2, 4, and 6 does take such extra terms, respectively. 

Regarding each group of figures (Figs. 1, 2; Figs. 3, 4; Figs. 5, 6), at slices corresponding to constant time in the range considered in the graphics, there is a subtle and prominent difference between the  warped horizons, regarding the respective corresponding eras. In Figs. 1 and 2 that regard a brane dominated by a cosmological constant,  the warped horizon of the associated McVittie black string increases monotonically along the extra dimension, irrespectively of the time and  independently whether the brane tension is constant or the brane tension is variable. 
In all other cases, it does not happen.
%If the brane is rigid

Notwithstanding, Figs. 3-6 evince another type of  behavior. Fig. 3 concerns a radiation-dominated FRW brane for a constant brane tension. The black string warped horizon is still a monotonic function along the extra dimension.  Instead, when one regards this kind of brane, by considering the variable brane tension, the graphic on the left  in Fig. 4 shows that for the time scale $t \lesssim 0.59$ (in the normalized scale used) the black string warped horizon decreases along the extra dimension, and for $t \gtrsim t_1 = 0.59$ it monotonically increases. For the graphic on the right in Fig. 4, that takes into account the extra terms in Eqs.(\ref{magnetico}, \ref{magnetico1}) regarding the variable brane tension, the effects due to those terms are even more drastic. For $y \gtrsim y_1 \sim  0.02$ the square of the black string horizon is negative, preventing the black string to exist for values greater than $y_1$. It means that the black string has a pancake-like shape \cite{maartens} and ceases to exist along the extra dimension for $y > y_1$.  It is a prominent result for the formalism here employed, based on the Taylor expansion (\ref{eletrico}): such procedure near the brane, in the range where the Taylor expansion holds, provides in this context all the information about the black string, and it is not solely a perturbative method.

For the matter-dominated FRW brane, the analysis is similar, although the same black string behavior happen later in time: in this situation $t_1 \sim 0.76$. Besides, along the extra dimension $y_1 \sim 0.018$. All features are similar as analyzed in the previous paragraph.

Indeed, as the graphics in the right hand side in Figs. 2, 4, and 6 are the most realistic ones, regarding respectively the cases where the Taylor expansion (\ref{eletrico}) is considered with the extra terms (\ref{magnetico}) and (\ref{magnetico1}) %
due to the variable brane tension, we illustrate below the McVittie black strings. As the graphics in Figs. 4 and 6 have a similar pattern, we depict2 below the black strings for $t=0.6$, respectively for the Fig. 2 and for Fig. 6 (similar to Fig. 4):
\begin{figure}[h]
\begin{minipage}{14pc}
\includegraphics[width=16pc]{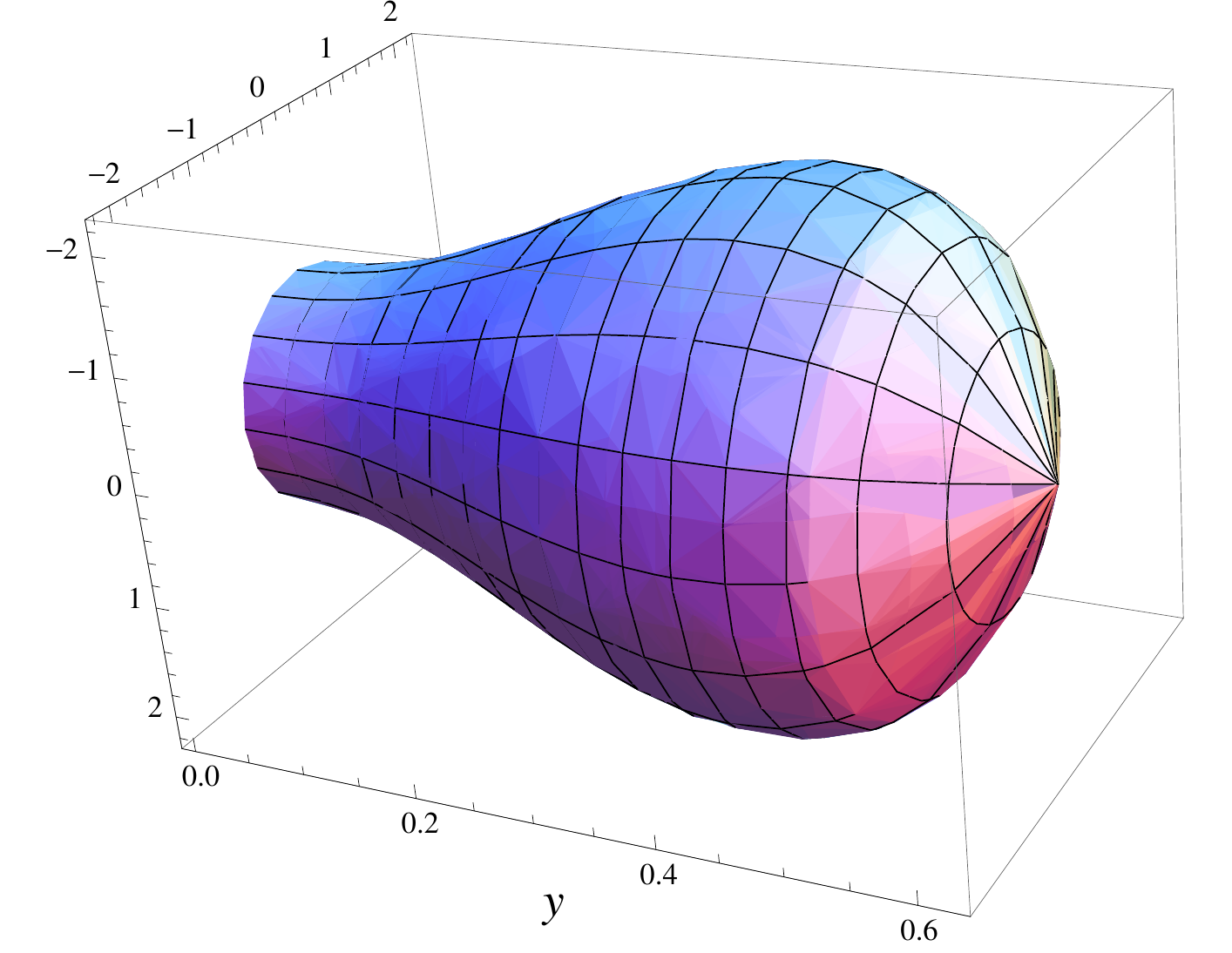}
\caption{\label{bs9} \footnotesize\; { Graphic of the black string 
for a Universe dominated by relativistic matter or radiation, where $a(t) \propto t^{\beta/2}$.}}
\end{minipage}\hspace{7pc}%
\begin{minipage}{14pc}
\includegraphics[width=16pc]{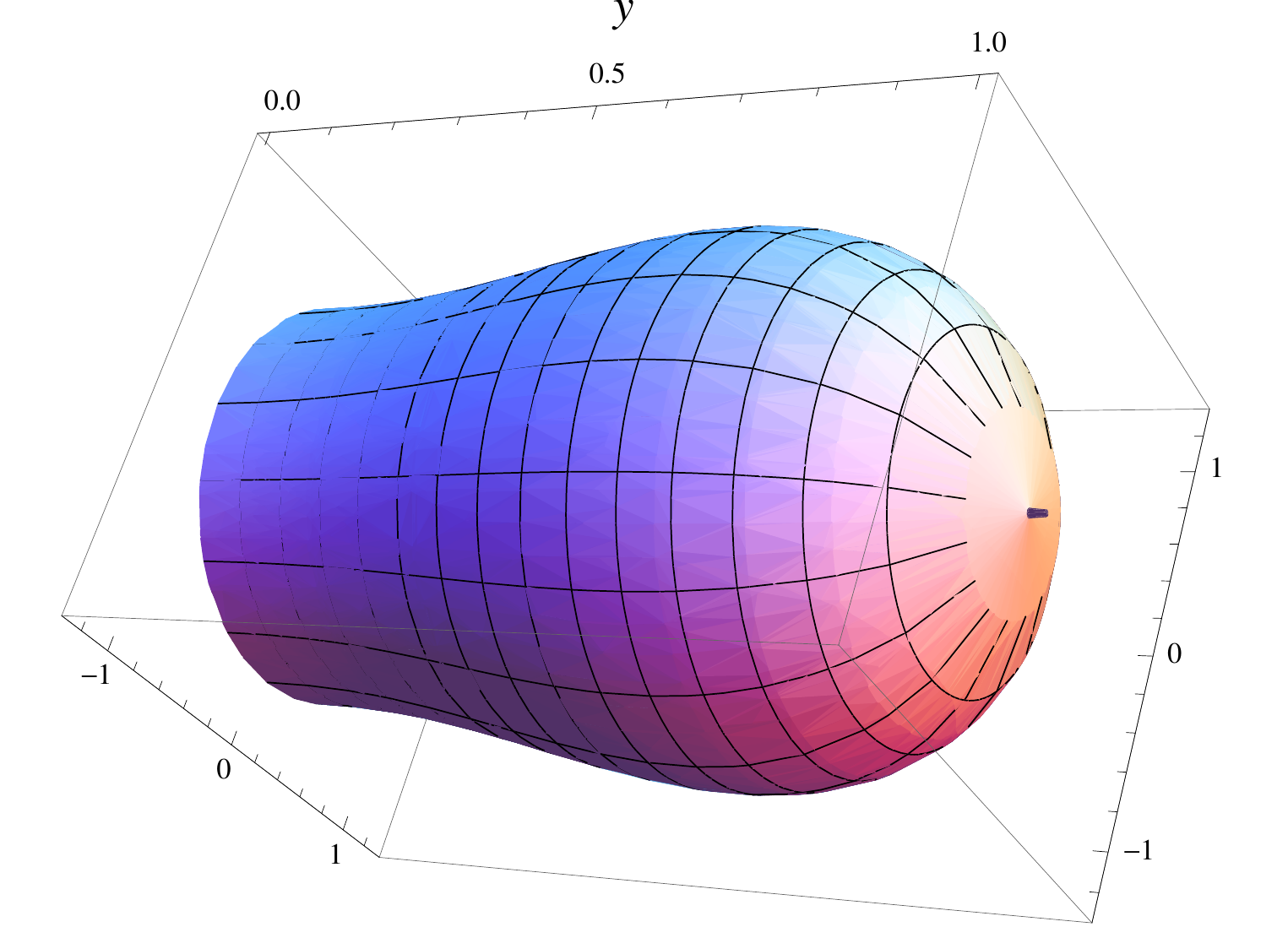}
\caption{\label{bs11} {Graphic of the black string 
for a Universe dominated by non-relativistic matter, where $a(t) \propto t^{2\beta/3}$.
}}\end{minipage}
\end{figure}
\begin{figure}[H]
\begin{center}\includegraphics[width=2.7in]{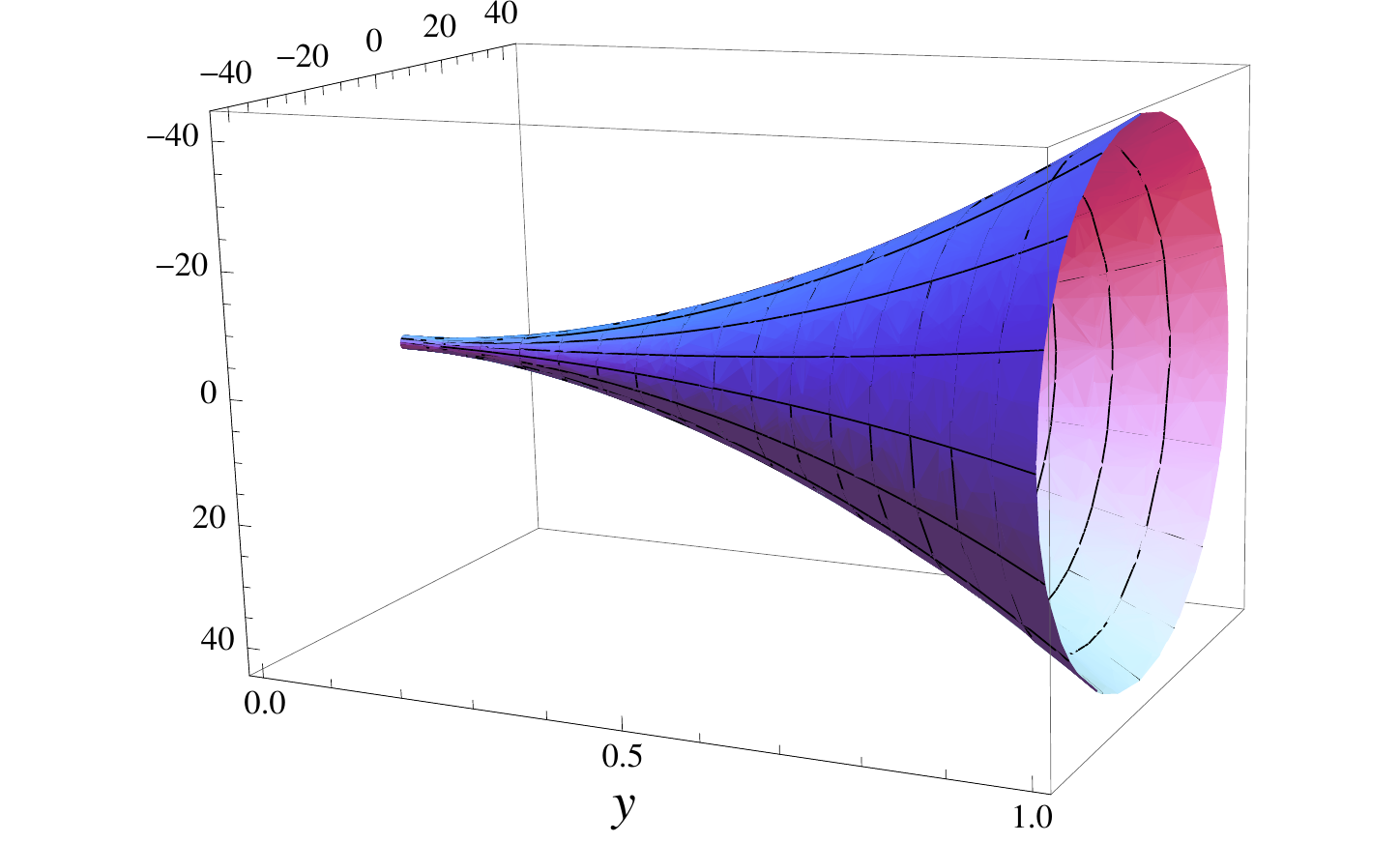}
\caption{\label{bs10}Graphic of the black string 
for a Universe dominated by relativistic matter or radiation, where $a(t) \propto \exp(H_0 t)$.}
\end{center}
\end{figure}
\noindent As already observed, the variable brane tension scenario brings drastic changes in the black string profiles, as depicted in Figs. \ref{bs9}-\ref{bs10}. In Fig. \ref{bs10} above on the left, the black string warped horizon always increases in full compliance to Fig. 2, for the case of a brane dominated by a cosmological constant. On the other hand, Figs. \ref{bs9} and \ref{bs11} show that there is a point along the extra dimension where the black string horizon tends to zero. It is an unexpected property, and in  Section IV we shall show, by analyzing the 5D Kretschmann invariants, that the vanishing of the black string horizon 
corresponds to a regular solution in the bulk, absent of physical singularities.

It is an immediate consequence of out analysis  that the brane is allowed to fluctuate in the variable tension paradigm. A possible interpretation may be given in terms of the emission and/or absorption of gravitons into the bulk implying the transference of momentum to the brane, being interpreted as a local deformation of the brane shape. In fact, combining the fact that a completely rigid object cannot exist in the general relativity framework with the presence of a scalar field representing the
brane position into the bulk, one arrives at the possibility of a
spontaneous symmetry breaking of the bulk diffeomorphism. In this
way, a perturbative spectrum of scalar particles, the so called
branons, may appear if the tension scale is much smaller than the
higher dimensional mass scale \cite{BRANON}.

In all cases for the cosmological evolution on the brane, described by (\ref{tensioneh0t}), (\ref{tensiont2b}), and (\ref{tension32b}), respectively corresponding to a Universe dominated by non-relativistic matter,  dominated
by the radiation or relativistic matter, and having a cosmological constant, the variable brane tension tends to a constant value as the time goes to infinity, and the thus brane becomes rigid.   

\section{Bulk Metric, the Black String and Variable Brane Tension: Removing Physical Singularities}

In this Section we aim to show that although the previous analysis on the black string profile regards a perturbative method, the determination about the bulk singularities consists of an exact method. The analysis of the 4D and 5D Kretschmann invariants evinces the character of the bulk solutions, that can be regular in  the whole bulk, not solely near the brane. Furthermore, we shall prove that for some eras of the evolution of the Universe, the singularities in the bulk are removed as the cosmological time elapses, due to the variable brane tension.

Actually, in the case where at late times the cosmology is dominated by a positive cosmological constant, the  metric (\ref{mcvitt}) on the brane is regular everywhere on and outside the associated black hole horizon, and it asymptotes to the Schwarzschild-de Sitter geometry, which has a Kottler black string   associated to it \cite{EPJC}.  When the cosmological constant equals zero our results are led to the ones in  \cite{Chamblin:1999by}. 
When $M=0$, the solution is led to a  homogeneous and isotropic FRW cosmology on the brane. For $H(t)=H_0$, the classical (Schwarzschild) black string \cite{maartens, Chamblin:1999by, meuhoff} or a Schwarzschild-de Sitter (or Kottler) black string of mass $M$  \cite{EPJC} is obtained. All curvature invariants on the null surface  equal their values on the horizon of a Schwarzschild-de Sitter generalized black string of mass $M$ \cite{EPJC} and positive Hubble constant. At least in the case  when $H_{0}>0$ the McVittie metric on the brane induces a black string \cite{plb2013}.

An alternative radial coordinate is defined \cite{nolann} as
${\tt r} = (1+\mu)^2 a(t) {\rm r}$, the McVittie metric (\ref{mcvitt}) reads
\begin{equation}
d s^2 = -g\; d t^2 - {2H\,{\tt r}}\,f^{-1/2}d {\tt r}\,d t + f^{-1}{d {\tt r}^2}  + {\tt r}^2 d\Omega_2,
\label{fin}\nonumber
\end{equation}\noindent where 
$f = 1-2M/{\tt r}$. 
On the brane, a null apparent horizon is placed at ${\tt r}={\tt r}_-$, which is the smaller positive root of $g({\tt r})=1- 2M/{\tt r} - H^2 {\tt r}^2=0$. 
 When $H$ equals a constant, the metric above is the Schwarzschild-de Sitter metric.  
 
 When ${\tt r}=2M$ and $t$ is finite, the McVittie solution has a curvature singularity at $\mu = 1$, {as the Ricci scalar has the form $R = 12H^{2}+ \frac{6}{\beta} \dot H\left(1-\frac{2M}{\tt r}\right)^{-1}$. 
The metric (\ref{fin}) evinces that in such case there is a spacelike 3-surface on the brane, as {\tt r} is fixed to $2M$, and consequently $d{\tt r} = 0$. Besides, the metric term in $dt^2$ is $g({\tt r}) = 4M^2H^2(t) > 0$, and the sphere has finite radius ${\tt r} = 2M$. This surface lies in the causal past of all spacetime points in the patch of the metric considered \cite{Kaloper:2010ec} and the black hole singularity is indeed localized near the brane.}
 
{The invariant  \[ \xi=(\nabla_{\mu} \nabla_{\nu} R_{\phi\psi\rho \sigma})(\nabla^{\mu} \nabla^{\nu} R^{\phi\psi\rho \sigma})\]  (here $\nabla_\mu$ denotes the covariant derivative on the brane) diverges at the horizon along ingoing null geodesics \cite{Kaloper:2010ec}.} This curvature invariant is very soft since it takes invariants involving at least two derivatives of the curvature to detect it. This 4D Kretschmann invariant can be related to its 5D counterpart, as the 5D and the 4D Riemann tensors are related by the Gauss equation as
$
{}^{(5)}R_{\phi\kappa\rho\sigma} = R_{\phi\kappa\rho\sigma}
 -K_{\phi\rho}K_{\kappa\sigma} + K_{\phi\sigma}K_{\kappa\rho}$. Consequently,
 the 5D version of the invariant $\xi$  
 reads
 \beq
 {}^{(5)}\xi=(D_a D_b {}^{(5)}R_{\phi\kappa\zeta \sigma})(D^aD^b {}^{(5)}R^{\phi\kappa\zeta\sigma}),\label{xi5}\eeq where $D_a$ denotes the 5D covariant derivative. {It is worthwhile to point that  $a,b$ are effectively  4D spacetime indexes, as the 5D covariant derivative can be realized as $D_a = \nabla_\mu$, for $a = 0,\ldots, 3$, and $D_a = \nabla_y$, when $a=5$. }The invariant (\ref{xi5}) was  shown to diverge at the black string warped horizon as well as in the McVittie black string singularity \cite{plb2013}, agreeing to the limit $a(t)=1$, corresponding to the classical black string. {Therefore, the difference between the 4D and 5D invariants is given by
 \begin{eqnarray}
{}^{(5)}\xi - \xi\!&=&\!
 2(\nabla_{\mu} \nabla_{\nu} K_{\tau[\rho\vert}K_{\psi\vert\sigma]})(\nabla^{\mu} \nabla^{\nu}K^{\tau\rho}K^{\psi\sigma})
-2 (\nabla_{y} \nabla_{\nu} K_{\tau\rho}K_{\psi\sigma})
(\nabla^{y} \nabla^{\nu}K^{\tau\sigma}K^{\psi\rho})
\nonumber\\&&+(\nabla_{(y} \nabla_{\nu)} R_{\tau\psi\rho \sigma})(\nabla^y \nabla^{\nu} R^{\tau\psi\rho \sigma})
-2(\nabla_{(\mu} \nabla_{y)} K_{\tau\rho}K_{\psi\sigma})(\nabla^{(\mu} \nabla^{y)} R^{\tau\psi\rho \sigma})
\nonumber
 \\&&- 
4 (\nabla_{\mu} \nabla_{y} K_{\tau[\rho\vert}K_{\psi\vert\sigma]})
(\nabla^{\mu} \nabla^{y}K^{\tau\sigma}K^{\psi\rho})+ (\nabla_y^2 R_{\tau\psi\rho \sigma})((\nabla^y)^2 R^{\tau\psi\rho \sigma})\nonumber\\&&-4(\nabla_y^2 K_{\tau\rho}K_{\psi\sigma})((\nabla^y)^2 R^{\tau\psi\rho \sigma})+2(\nabla_y^2 K_{\tau[\sigma\vert}K_{\psi\vert\rho]})((\nabla^y)^2 K^{\tau\sigma}K^{\psi\rho})\,.\label{xi66}\end{eqnarray} By considering the extrinsic curvature in (\ref{kurvmc}) for the McVittie solution, we can calculate
 ${}^{(5)}\xi$. 
   
Let us first write the 4D Kretschmann scalar $\xi$ for the McVittie solution (\ref{mcvitt}): 
 \ba
\xi&=& \frac{1}{{\tt r}^{17} \!\left(1\!-\!\frac{2M}{{\tt r}}\right)^{\frac{11}{2}}}\Biggl[12\sqrt{1\!-\!\frac{2 M}{{\tt r}}} {\tt r}^{13}{H}^4 \Bigl( (885 M^4\!-\!1320 M^2 (2 M\!-\!{\tt r})^5\!-\!1686 M^3 {\tt r}\!+\!1240 M^2 {\tt r}^2  \nonumber \\
&&-412 M {\tt r}^3 +52 {\tt r}^4 )    {\dot H}^2 \Bigr)\!+\!24 {\tt r}^{13} \left(158 M^4\!-\!185 M^3 {\tt r}\!+\!63 M^2 {\tt r}^2\!-\!M {\tt r}^3\!-\!2 {\tt r}^4\right)  {H}^3   {\dot H}{\ddot  H} \nonumber\\ && -24 (2 M\!-\!{\tt r}) {\tt r}^{10}  {H}    {\ddot H}  \biggl(M^2 \!\left(6 M^2\!\!-\!\!19 M {\tt r}\!+\!8 {\tt r}^2\right)\!{\dot H}  +\sqrt{1\!-\!\frac{2 M}{{\tt r}}} {\tt r}^4  \left(67 M^2\!-\!56 M {\tt r}\!+\!12 {\tt r}^2\right) \!  {\dot H} ^2 \nonumber \\
&&+  {\tt r}^4 \left(14 M^2-11 M {\tt r}+2 {\tt r}^2\right)  \dddot{H} \biggr)   \!+\! 4 {\tt r}^3\! (-2 M+{\tt r})  {H} ^2 \biggl[6 M {\tt r}^{10} \left(47 M {\tt r}\!-\!57 M^2\!-\!10 {\tt r}^2\right)  {\dot H} ^3\nonumber  \\  
 &&
-2 M^2 \sqrt{1\!-\!\frac{2 M}{{\tt r}}} {\tt r}^7 \left(334 M^2\!-\!347 M {\tt r}\!+\!96 {\tt r}^2\right)    {\dot H} ^2
\! +\!3 \sqrt{1\!-\!\frac{2 M}{{\tt r}}} \Bigl(240 M^2 (25 M\!-\!12 {\tt r}) {\tt r}^4\!-\!2M \nonumber  \\
 && +{\tt r}^{11}\! \left(109 M^2\!-\!88 M {\tt r}\!+\!
 19 {\tt r}^2\right){\ddot H} ^2\Bigr)
  \!-\!6 M {\tt r}^3{\dot H}\Bigl(180 M {\tt r}^4\!-\!2M\sqrt{1\!-\!\frac{2 M}{{\tt r}}} {\tt r}^8 (-5 M+{\tt r})  { \dddot{H}} \Bigr)\biggr] \nonumber
 \\  
 && +4 ({\tt r}\!-\!2M) \biggl[ 2 M^2 {\tt r}^{10} \left(6 M^2+M {\tt r}-2 {\tt r}^2\right)   {\dot H} ^3+3 \sqrt{1\!-\!\frac{2 M}{{\tt r}}} {\tt r}^{14} \left(57 M^2\!-\!52 M {\tt r}\!+\!12 {\tt r}^2\right)   {\dot H} ^4 \nonumber
 \\&&
 +{\tt r}^7 ({\tt r}\!-\!2M)   {\dot H} ^2 \left(M^2 \sqrt{1\!-\!\frac{2 M}{{\tt r}}} \left(847 M^2-832 M {\tt r}+222 {\tt r}^2\right)-6 (5 M-2 {\tt r}) {\tt r}^7  \dddot{H} \right)
\nonumber  \\ &&
 -2 M^2 (2 M-{\tt r}) {\tt r}^3   {\dot H}  \left(180 M (2 M-{\tt r})^3+7 \sqrt{1\!-\!\frac{2 M}{{\tt r}}} {\tt r}^8  \dddot{H} \right)
\nonumber \\&&
 +3{\tt r}\!\left({1-\frac{2 M}{{\tt r}}}\right)^{3/2}\!\!\!\!\bigl(120 M^2 ({\tt r}\!-\!2M)^3 \left(65 M^2\!\!-\!60 M {\tt r}\!+\!14 {\tt r}^2\right)  \!+\!{\tt r}^{15}  \dddot{H} ^2\bigr)\!-\!\frac{40}{3} M^2 {\tt r}^{11}   {\ddot H} ^2\biggr] \Biggr]\,
\label{krettxi}
\ea\noindent This 4D Kretschmann invariant  encodes the existence of two physical singularities on the brane, at ${\tt r} = 0$ and ${\tt r} = 2M$. Such kind of singularity is soft, as it involves at least two derivatives of the curvature to detect it. Besides, the singularity at the surface ${\tt r}=2M$ is a soft, null naked singularity when $t\rightarrow\infty$ in an FRW brane if the  Hubble parameter $H(t)$ goes to zero at late times \cite{Kaloper:2010ec}, which is the case  when $a(t) \propto t^{\beta/2}$ and $a(t) \propto t^{2\beta/3}$.

In order to calculate the 5D Kretschmann invariant (\ref{xi5}), all the terms at the right-hand side in (\ref{xi66}) are computed. As the expression for ${}^{(5)}\xi$ is extensive, all the information about it and the associated singularities is depicted in the graphics in what follows, where we used the normalization $M=1$ for the sake of simplicity, without loss of generality. In what follows we depict the graphics of the 5D Kretschmann scalar for all the eras of the evolution of the Universe. We  show that the vanishing of the black string warped horizon, for a Universe dominated by matter or radiation (see Figs. \ref{bs11} and \ref{bs10}) is accompanied with the banishment of the physical singularities in the bulk.

\newpage
\begin{figure}[h]
\begin{minipage}{14pc}
\includegraphics[width=17pc]{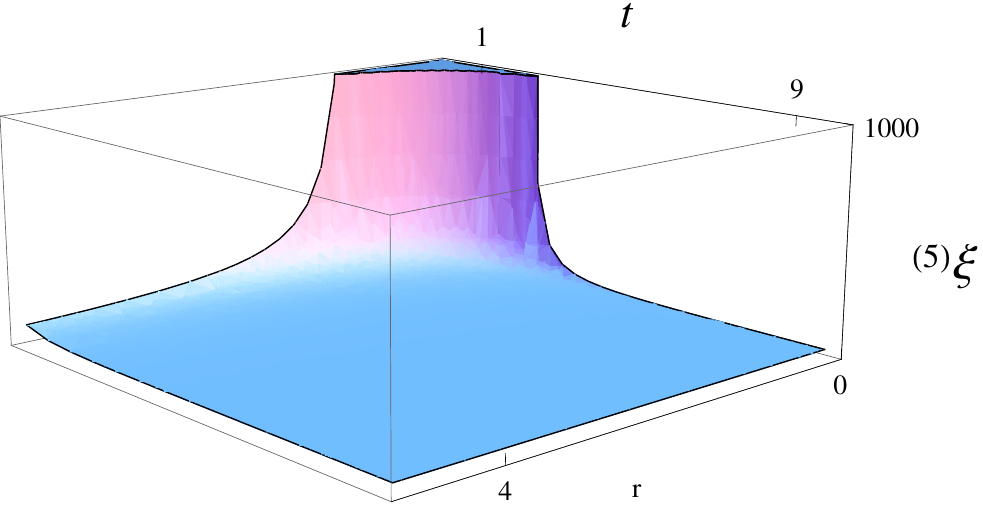}
\caption{\label{kretd} \footnotesize\; {Plot  of the 5D Kretschmann scalar ${}^{(5)}\xi$ 
as a function of time and ${\tt r}$, for the scale factor $a(t) \propto t^{\beta/2}$.}}
\end{minipage}\hspace{7pc}%
\begin{minipage}{14pc}
\includegraphics[width=17pc]{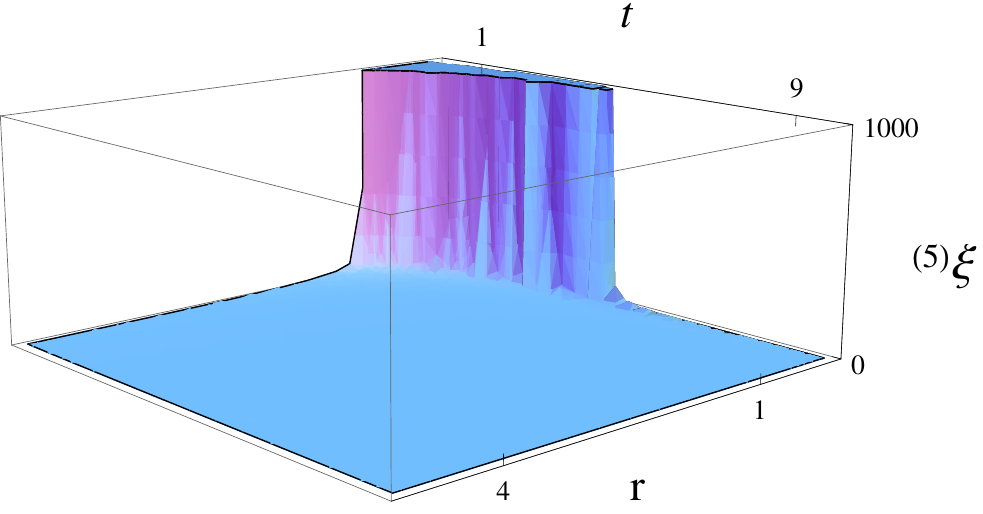}
\caption{\label{kret5d2} {Plot  of the 5D Kretschmann scalar ${}^{(5)}\xi$ 
as a function of time and ${\tt r}$, for the scale factor $a(t) \propto t^{2\beta/3}$.}}\end{minipage}
\end{figure}
Figs. \ref{kretd} and \ref{kret5d2} reveal that 
as time elapses, the singularities in the brane are removed, providing a regular 5D bulk solution, as the 5D Kretschmann invariants do not diverge for both cases of a Universe dominated by radiation or matter.
 
Fig. \ref{kretdexp} evinces that two physical singularities on the brane at ${\tt r} = 0$ and ${\tt r} = 2M$ remain in the bulk along the extra dimension and there is no additional singularity in the bulk, in a Universe dominated by a cosmological constant.
 
 \begin{figure}[h]
\begin{minipage}{14pc}
\includegraphics[width=17pc]{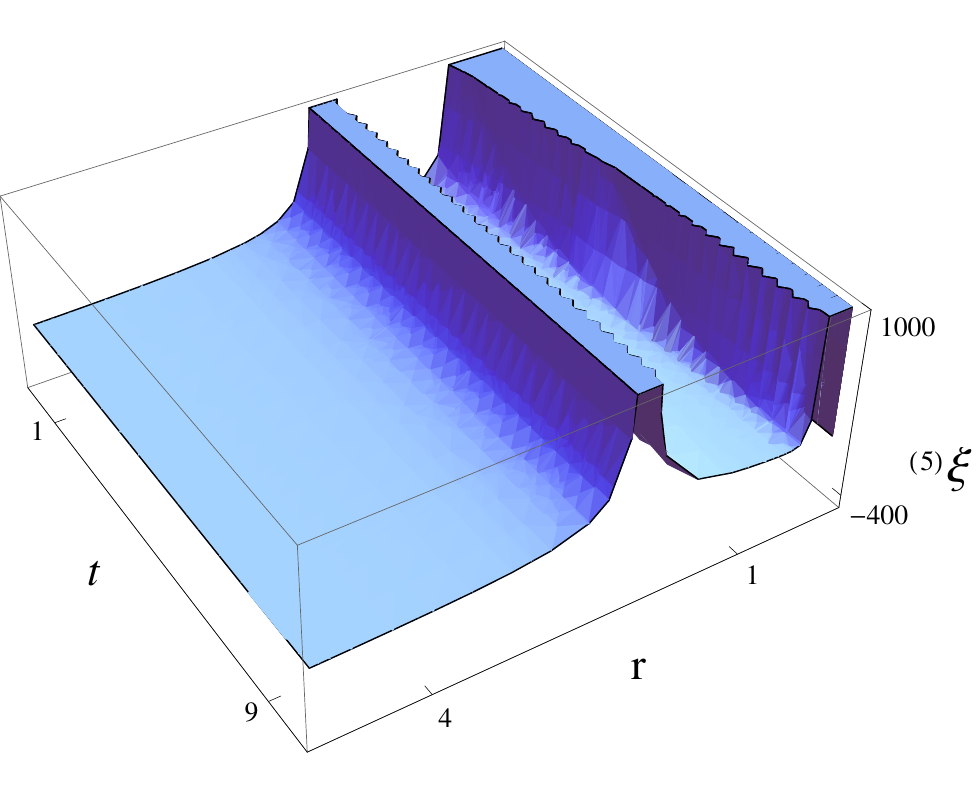}
\caption{\label{kretdexp} \footnotesize\; {Plot  of the 5D Kretschmann scalar ${}^{(5)}\xi$ 
as a function of time and ${\tt r}$, for the scale factor $a(t) \propto \exp(H_0 t)$.}}
\end{minipage}\hspace{7pc}%
  \end{figure}
}

\section{Concluding Remarks}
Since no single perfect fluid description can be used as a source for the generalized McVittie solution, to find a suitable single-fluid interpretation for the metric it is  required the introduction of viscosity and heat transport as well, and the analysis here can be accomplished in the context of the gravity/fluid correspondence \cite{navier}. 
Hence, a single imperfect fluid can be used as a source to obtain a generalized McVittie metric as an exact solution to Einstein equations, and that the mass variation can be interpreted as a consequence of heat flow in the radial direction within the fluid. An accreting black hole model was used in \cite{abd} to unravel its differences with respect to the static-mass case, keeping the necessary conditions for the McVittie metric to be interpreted as a black hole at future infinity.  A generalized black string can be obtained in this sense. 

In our approach, the additional terms  in the black string warped horizon (\ref{mag2}) metric are shown to provide modifications in the McVittie black string warped horizon in a variable tension braneworld scenario.
The black string associated to the McVittie solution of the Einstein field equations is shown to be drastically modified by the terms
due to the Universe expansion.  In particular,  the well known results in the literature are recovered as limiting cases
when $a(t)=1$ or $M=0$. 
For radiation-dominated and matter-dominated FRW branes,  the Taylor expansion provides an exact method, providing all the information about the black string warped horizon along the extra dimension, and it is not a mere  perturbative method. When the variable brane tension is taken into account 
there is a value for the time coordinate beyond which the black string warped horizon is zero thereupon, what does mean that the black string ceases to exist along the extra dimension. {As illustrated in Figs. 7, the McVittie black string warped horizon 
can have a completely different profile in a variable brane tension scenario.
The black string warped horizon along the extra dimension provides immediate information on the black string stability under small perturbations, as the (Schwarzschild)  black string Gregory-Laflamme instability \cite{greg}. Indeed, the horizon $\sqrt{g_{\theta\theta}({\tt r}, t, y)}$ can collapse to zero before the perturbation takes part, as illustrated in the Figs. 3-6, for adequate ranges in the variable $t$ therein. The determination whether the McVittie black string is unstable or not, under Gregory-Laflamme perturbations, is out of the scope of this work.

By analyzing the 4D and 5D Kretschmann invariants, we showed that  the black string warped horizon vanishes along  the extra dimension, for a Universe dominated by matter or radiation, what induces  the bulk singularities to disappear, leaving a regular
bulk solution. Moreover, no additional singularity is introduced in the bulk, with respect to  the brane  black hole
physical singularities, when the expanding Universe is dominated by a cosmological constant. The analysis on how the black string warped horizon leaks into the bulk near the brane concerned a perturbative method, widely used in the literature. Notwithstanding, the analysis of the bulk singularities relies on an exact method provided by the Gauss equation and Eqs.(\ref{xi5}) and (\ref{xi66}). Therefore, the 4D and 5D Kretschmann invariants show that the bulk solutions,  can be regular in  the whole bulk, and in  some eras of the evolution of the Universe, the singularities in the bulk are removed as the cosmological time elapses, due to the variable brane tension.
}

Branons and the brane flexibility  \cite{branons, branons1, branons3} are related and  some cosmological and astrophysical constraints on the brane tension were considered in \cite{branons5}. There are bounds on the brane tension and on the branon mass, in the case where the brane tension scale is much smaller than  the 5D fundamental scale of gravity \cite{branons5}, and the first indications of extra dimensions may arise by the production of branons, {allowing to measure the brane tension \cite{alca}, and consequently the feasibility concerning the more general assumption of a variable brane tension}. 
In the context of a variable brane tension, according to this interpretation, the contribution to the branons creation 
may be notorious, as well as its influence on the
black string warped horizon profile. 
\section*{Acknowledgments}
D. Bazeia would like to thank CAPES, CNPq and FAPESP for financial support. R. da Rocha is grateful to CNPq grant 303027/2012-6 and  480482/2012-8. J. M. Hoff da Silva thanks to CNPq for partial financial support.


\begin{thebibliography}{99}
\footnotesize



\bibitem{ran} L. Randall and R. Sundrum, \emph{A Large Mass Hierarchy from a Small Extra Dimension, Phys. Rev. Lett.} {\bf 83} (1999) 3370 [{\tt arXiv:hep-ph/9905221}].

\bibitem{maar2000}
  R.~Maartens,
  \emph{Cosmological dynamics on the brane,
  Phys.\ Rev.\ D} {\bf 62} (2000) 084023
  [{\tt arXiv:hep-th/0004166}].

\bibitem{binetruy}
P.~Binetruy, C.~Deffayet, U.~Ellwanger, D.~Langlois,
  \emph{Brane cosmological evolution in a bulk with cosmological constant},
  Phys.\ Lett.\ B {\bf 477} (2000) 285
  [{\tt arXiv:hep-th/9910219}].
  \bibitem{bazeia1}
  D.~Bazeia, F.~A.~Brito, F.~G.~Costa,
  \emph{First-order framework and domain-wall/brane-cosmology correspondence},
  Phys.\ Lett.\ B {\bf 661} (2008) 179
  [{\tt arXiv:0707.0680 [hep-th]}].

\bibitem{gly1} L. A. Gergely, \emph{Friedmann branes with variable tension, Phys. Rev. D} {\bf  78} (2008) 084006 [{\tt arXiv: 0806.3857 [gr-qc]}].

\bibitem{gly2} L. A. Gergely, \emph{E\"otv\"os branes, Phys. Rev. D} {\bf  79} (2009) 086007 [{\tt arXiv: 0806.4006 [gr-qc]}].

\bibitem{bulk1} M. C. B. Abdalla, J. M. Hoff da Silva, R. da Rocha, \emph{Notes on the Two-brane Model with Variable Tension, 
Phys. Rev. D} {\bf 80} (2009) 046003 [{\tt arXiv: 1101.4214 [gr-qc]}].

\bibitem{bulk2} J. M. Hoff da Silva, \emph{Two-branes with variable tension model and the effective Newtonian constant, 
Phys. Rev. D} {\bf 83} (2011) 066001 [{\tt arXiv:0907.1321 [hep-th]}].


\bibitem{european}  K. C. Wong, K. S. Cheng, T. Harko, \emph{Inflation and late time acceleration in braneworld cosmological models with varying brane tension, 
Eur. Phys. J. C} {\bf 68} (2010) 241 {[\tt arXiv:1005.3101 [gr-qc]}].


\bibitem{branons}  M.~Bando, T.~Kugo, T.~Noguchi, K.~Yoshioka,
\emph{Brane fluctuation and suppression of Kaluza-Klein mode couplings,
  Phys.\ Rev.\ Lett.}\  {\bf 83} (1999) 3601
  [{\tt arXiv:hep-ph/9906549}]. 

  
  \bibitem{emp2}
  R.~Emparan, A.~Fabbri, N.~Kaloper,
  \emph{Quantum black holes as holograms in AdS brane worlds},
  \emph{JHEP} {\bf 0208} (2002) 043
  [{\tt arXiv:hep-th/0206155}]. 
 \bibitem{emp3}  T.~Tanaka,
  \emph{Classical black hole evaporation in Randall}-\emph{Sundrum infinite brane world,
  Prog.\ Theor.\ Phys.\ Suppl.}\  {\bf 148} (2003) 307
  [{\tt arXiv:gr-qc/0203082}].
  

  
 
   
  
  
  
\bibitem{maartens} R.~Maartens, K. Koyama, {\it Brane world gravity},
   \emph{Living Rev. Relativity} {\bf 13} (2010) 5  [{\tt arXiv:gr-qc/0312059}].
   
   
  \bibitem{yoshino}
  H.~Yoshino,
  \emph{On the existence of a static black hole on a brane},
  \emph{JHEP} {\bf 0901} (2009) 068
  [{\tt arXiv:0812.0465}].
   

   
 

\bibitem{Gergely:2006hd}
  L.~A.~Gergely, \emph{Black holes and dark energy from gravitational collapse on the brane,}
  \emph{JCAP} {\bf 02 } (2007)  027
  [{\tt  arXiv:hep-th/0603254}].
  
  
\bibitem{Anderson:2005af}
  E.~Anderson, R.~Tavakol,
 \emph{Geodesics, the equivalence principle and singularities in higher-dimensional general relativity and braneworlds},
  \emph{JCAP} {\bf 10 } (2005)  017 
  [{\tt arXiv:gr-qc/0509055 [gr-qc]}].


  
\bibitem{mcvittie}
G.~C.~McVittie, \emph{The mass-particle in an expanding universe, Mon. Not. R. Astron. Soc.} {\bf 93} (1933) 325.

  \bibitem{Kaloper:2010ec}
  N.~Kaloper, M.~Kleban, D.~Martin,
  \emph{McVittie's Legacy: Black Holes in an Expanding Universe},
  \emph{Phys.\ Rev.\ D} {\bf 81} (2010) 104044
  [{\tt arXiv:1003.4777 [hep-th]}].
 
  
 \bibitem{plb2013} D.~Bazeia, J.~M.~Hoff~da Silva, R.~da Rocha,
\emph{Black holes in realistic branes: black string-like objects?,}  Phys. Lett. B {\bf 721} (2013) 306 [{\tt arXiv:1303.2243 [gr-qc]}].

  
  \bibitem{lake1}
  K.~Lake, M.~Abdelqader,
  \emph{More on McVittie's Legacy: A Schwarzschild - de Sitter black and white hole embedded in an asymptotically $\Lambda$CDM cosmology},
  \emph{Phys.\ Rev.\ D} {\bf 84} (2011) 044045
  [{\tt arXiv:1106.3666 [gr-qc]}].
  
 
  \bibitem{Carrera:2009ve}
  M.~Carrera and D.~Giulini,
  \emph{On the generalization of McVittie's model for an inhomogeneity in a cosmological spacetime,
  Phys.\ Rev.\ D} {\bf 81} (2010) 043521
  [arXiv:0908.3101 [gr-qc]].

\bibitem{faraoni}
  V.~Faraoni, A.~F.~Zambrano Moreno, R.~Nandra,
  \emph{Making sense of the bizarre behaviour of horizons in the McVittie spacetime},
  \emph{Phys.\ Rev.\ D} {\bf 85} (2012) 083526
  [{\tt arXiv:1202.0719 [gr-qc]}].


 \bibitem{nolann}
B.~C.~Nolan,
\emph{A point mass in an isotropic universe: Existence, uniqueness and basic
properties},
{\it Phys.\ Rev.\ D} {\bf 58},  064006 (1998);
[{\tt arXiv:gr-qc/9805041}].


\bibitem{abd}
  D.~C.~Guariento, M.~Fontanini, A.~M.~da Silva and E.~Abdalla,
  \emph{Realistic fluids as source for dynamically accreting black holes in a cosmological background},
  Phys.\ Rev.\ D {\bf 86} (2012) 124020
  [{\tt arXiv:1207.1086 [gr-qc]}].


  \bibitem{lake2}
  P.~Landry, M.~Abdelqader and K.~Lake,
  \emph{The McVittie solution with a negative cosmological constant,
  Phys.\ Rev.\ D} {\bf 86} (2012) 084002
  [{\tt arXiv:1207.6350 [gr-qc]}].
 
\bibitem{Ferraris:1996ey}
  M.~Ferraris, M.~Francaviglia and A.~Spallicci,
  \emph{Associated radius, energy and pressure of McVittie's metric, in its astrophysical application,
  Nuovo Cim.\ B} {\bf 111} (1996) 1031.
  \bibitem{Haines:1993sd}
  P.~Haines and J.~D.~Harris,
  \emph{Thin shells in flat McVittie space-times,
  Astrophys.\ J.}\  {\bf 418} (1993) 579.
  \bibitem{Patel:1999ej}
  L.~K.~Patel, R.~Tikekar and N.~Dadhich,
  \emph{Higher dimensional analog of McVittie solution},
  \emph{Grav.\ Cosmol.}\  {\bf 6} (2000) 335
  [{\tt arXiv:gr-qc/9909069}].

\bibitem{sakaihaines}
  N.~Sakai and P.~Haines,
  \emph{Peculiar Velocities of Nonlinear Structure: Voids in McVittie Spacetime,
Astrophys.\ J.} {\bf 536}, 515 (2000).  
[{\tt arXiv:astro-ph/9909183}].

\bibitem{meuhoff} R.{} da Rocha, J.{} M.{} Hoff da Silva, 
\emph{Black string corrections in variable tension braneworld scenarios,} 
\emph{Phys. Rev. D} {\bf 85} (2012) 046009 [{\tt arXiv:1202.1256 [gr-qc]}].

{\bibitem{greg}
  R.~Gregory and R.~Laflamme,
  \emph{Black strings and p-branes are unstable,}
  \emph{Phys.\ Rev.\ Lett.}\  {\bf 70} (1993) 2837
  [{\tt arXiv:hep-th/9301052}].}


{ \bibitem{Kanti1}
  P.~Kanti and K.~Tamvakis,
  \emph{Quest for localized 4-D black holes in brane worlds},
  Phys.\ Rev.\ D {\bf 65} (2002) 084010
  [{\tt arXiv:hep-th/0110298}].
  
  \bibitem{Kanti2}
  P.~Kanti, I.~Olasagasti and K.~Tamvakis,
 \emph{Quest for localized 4-D black holes in brane worlds. 2. Removing the bulk singularities},
  Phys.\ Rev.\ D {\bf 68} (2003) 124001
  [{\tt arXiv:hep-th/0307201}].}
 \bibitem{3333} T. Shiromizu, K. Maeda, M. Sasaki, 
\emph{The Einstein Equations on the 3-Brane World, 
Phys. Rev. D} {\bf 62} (2000) 043523 [{\tt arXiv:gr-qc/9910076}]; A. N. Aliev  A. E. Gumrukcuoglu, {\it Gravitational Field Equations on and off a 3-Brane World, 
Class. Quant. Grav.} {\bf 21} (2004) 5081 [{\tt arXiv:hep-th/0407095}].


  \bibitem{casadio2004}
  R.~Casadio and C.~Germani,
  \emph{Gravitational collapse and black hole evolution: Do holographic black holes eventually 'anti-evaporate'?,}
  Prog.\ Theor.\ Phys.\  {\bf 114} (2005) 23
  [{\tt arXiv:hep-th/0407191}].

\bibitem{clark} S.~S.~Seahra, C.~Clarkson, R.~Maartens, \emph{Detecting extra dimensions with gravity wave spectroscopy: the black string brane-world, 
  Phys.\ Rev.\ Lett.}  {\bf 94} (2005) 121302
  [{\tt arXiv:gr-qc/0408032}].


%  \bibitem{Chamblin:1999by}
 % A.~Chamblin, S.~W.~Hawking and H.~S.~Reall,
 % \emph{Brane world black holes},
 % Phys.\ Rev.\ D {\bf 61} (2000) 065007
  %[{\tt arXiv:hep-th/9909205}].  
  
 
\bibitem{all}  N.~Arkani-Hamed and M.~Schmaltz,
  \emph{Hierarchies without symmetries from extra dimensions},
  \emph{Phys.\ Rev.\ D} {\bf 61} (2000) 033005
  [{\tt arXiv:hep-ph/9903417}]; J.~D.~Lykken, R.~C.~Myers and J.~Wang,
  \emph{Gravity in a box}, 
  \emph{JHEP} {\bf 09} (2000) 009
  [{\tt arXiv:hep-th/0006191}].


  \bibitem{Chamblin:1999by}
  A.~Chamblin, S.~W.~Hawking and H.~S.~Reall,
  \emph{Brane world black holes},
  \emph{Phys.\ Rev.\ D} {\bf 61} (2000) 065007
  [{\tt arXiv:hep-th/9909205}].  

\bibitem{buchdahl} H. A. Buchdahl, 
\emph{Isotropic coordinates and Schwarzschild metric, 
Int. J. Theor. Phys.} {\bf 24} (1985) 731.


\bibitem{EPJC}
  J.~M.~Hoff da Silva and R.~da Rocha,
  \emph{Schwarzschild generalized black hole horizon and the embedding space},
 \emph{ Eur.\ Phys.\ J.\ C} {\bf 72} (2012) 2258
  [{\tt arXiv:1212.2588 [gr-qc]}].


 
  \bibitem{Eotvos} R. E\"otv\"os, \emph{Ueber den Zusammenhang der Oberfl\"achenspannung der Fl\"ussigkeiten mit ihrem Molecularvolumen}, Annalen der  Physik  {\bf 263} (1886) 448.
\bibitem{WILL} C. M. Will, \emph{The confrontation between General Relativity and experiment, Living Rev. Relativity} {\bf 9} (2005) 3 [{\tt arXiv:gr-qc/0510072}].

\bibitem{BRANON}  J. A. R. Cembranos, A. Dobado, and A. L. Maroto, \emph{Brane-World dark matter, Phys. Rev. Lett.} {\bf 90} (2003) 241301 [{\tt arXiv:hep-ph/0302041}].

\bibitem{navier} S.~Bhattacharyya, V.~E. Hubeny, S.~Minwalla,  M.~Rangamani,
\emph{Nonlinear Fluid Dynamics from Gravity}, {\textit{JHEP} {\bf 02}
(2008) 045} 
[{{\tt arXiv:0712.2456
[hep-th]}}]; S.~Bhattacharyya, S.~Minwalla, and S.~R. Wadia, 
\emph{The Incompressible Non-Relativistic Navier-Stokes Equation from Gravity},
{\textit{JHEP}
\textbf{08} (2009) 059}
[{{\tt arXiv:0810.1545 [hep-th]}}]; I. Breedberg, C. Keller, V. Lysov, A. Strominger,  
\emph{From Navier-Stokes to Einstein} 
 [{\tt arXiv: 1101.2451 [hep-ph]}].



\bibitem{branons1}  G.~R.~Dvali, I.~I.~Kogan, M.~A.~Shifman, 
\emph{Topological effects in our brane world from extra dimensions,
  Phys.\ Rev.\ D} {\bf 62} (2000) 106001
  [{\tt arXiv:hep-th/0006213}]. 
  \bibitem{branons3}
  H.~Murayama, J.~D.~Wells, \emph{Graviton emission from a soft brane,
  Phys.\ Rev.\ D} {\bf 65} (2002) 056011
  [{\tt arXiv:hep-ph/0109004}].   \bibitem{branons5}
  T.~Kugo, K.~Yoshioka, \emph{Probing extra dimensions using Nambu-Goldstone bosons,
  Nucl.\ Phys.\ B} {\bf 594} (2001) 301
  [{\tt arXiv:hep-ph/9912496}]. 
  
  
  \bibitem{alca}
  J.~Alcaraz, J.~A.~R.~Cembranos, A.~Dobado, A.~L.~Maroto,
  \emph{Limits on the brane fluctuations mass and on the brane tension scale from electron positron colliders,
  Phys.\ Rev.\ D} {\bf 67} (2003) 075010
  [{\tt arXiv:hep-ph/0212269}].
  


\end{thebibliography}
\end{document}